\documentclass[smallextended]{svjour3}       

\smartqed  
\usepackage{graphicx}
\usepackage[flushleft]{threeparttable}
\usepackage{amsmath,amsfonts,amssymb}
\usepackage{graphicx}
\usepackage{color}
\usepackage{array,multirow}
\usepackage[colorlinks=true,linkcolor=blue,citecolor=blue]{hyperref}

\newcommand{\saopaulo}{{S\~{a}o Paulo}}
\newcommand{\hi}{H{\sc i}}

\begin{document}

\title{Baryon acoustic oscillations from Integrated Neutral Gas Observations:\\ Broadband corrugated horn construction and testing}


\titlerunning{BINGO Broadband corrugated horn construction and testing}        

\author{C.~A.~Wuensche$^{1*}$ \and 
	 L.~Reitano$^{1}$ \and 
	 M.~W.~Peel$^{2,3,4}$ \and
	 I.~W.~A.~Browne$^{5}$ \and
	 B.~Maffei$^{6}$ \and
	 E.~Abdalla$^{2}$ \and 
	 C.~Radcliffe$^{7}$ \and
     F. Abdalla$^{8}$ \and
	 L. Barosi$^{9}$ \and
	 V.~Liccardo$^{1}$ \and
	 E.~Mericia$^{1}$ \and
	 G. Pisano$^{10}$ \and
	 C.~Strauss$^{1}$ \and
	 F.~Vieira$^{1}$ \and
	 T.~Villela$^{1,11}$ \and
	 B. Wang$^{12}$
}

\institute{C.~A.~Wuensche (orcid.org/0000-0003-1373-4719) \at
              \email{ca.wuensche@inpe.br} 
              \and
$^{1}$Divis\~{a}o de Astrof\'{\i}sica, Instituto Nacional de Pesquisas Espaciais (INPE), S\~{a}o Jos\'e dos Campos, SP, Brazil  \\
${^2}$Instituto de F\'{i}sica, Universidade de \saopaulo, \saopaulo~- SP, Brazil\\
$^{3}$Instituto de Astrof\'{i}sica de Canarias, E-38205 La Laguna, Tenerife, Spain\\
$^{4}$Departamento de Astrof\'{i}sica, Universidad de La Laguna (ULL), E-38206 La Laguna, Tenerife, Spain\\
$^{5}$Jodrell Bank Centre for Astrophysics, Alan Turing Building, Department of Physics and Astronomy, School of Natural Sciences, The University of Manchester, Oxford Road, Manchester, M13 9PL, UK\\
$^{6}$Institut d'Astrophysique Spatiale, Universit\'e Paris-Sud / Paris-Saclay, 91405 Orsay Cedex, France\\
$^{7}$Phase2 Microwave Ltd., Unit 1a, Boulton Rd, Pin Green Ind. Est., Stevenage, SG1 4QX, UK \\
$^{8}$Department of Physics \& Astronomy, University College London, Gower Place, London WC1E 6BT, UK \\
$^{9}$Unidade Acad\^emica de F\'{i}sica, Universidade Federal de Campina Grande, Campina Grande, PB, Brazil \\
$^{10}$School of Physics and Astronomy, Cardiff University, Cardiff, CF10 3AT, UK\\
$^{11}$Instituto de F\'{i}sica, Universidade de Bras\'{i}lia, Bras\'{i}lia, DF, BraZil\\
$^{12}$Center for Gravitation and Cosmology, YangZhou University, Yangzhou 225009, China \\
}

\date{Received: date / Accepted: date}

\maketitle

\begin{abstract}
The Baryon acoustic oscillations from Integrated Neutral Gas Observations (BINGO) telescope is a \mbox{40-m~class} radio telescope under construction that has been designed to measure the large-angular-scale intensity of \hi\ emission at 980--1260\,MHz and hence to constrain dark energy parameters. A large focal plane array comprising of 1.7-metre diameter, 4.3-metre length corrugated feed horns is required  in order to optimally illuminate the telescope. Additionally, very clean  beams with low sidelobes across a broad frequency range are required, in order to facilitate the separation of the faint \hi\ emission from bright Galactic foreground emission. Using novel construction methods, a  full-sized prototype horn has been assembled. It has an average insertion loss of around $-0.15$\,dB across the band, with a return loss around $25$\,dB. The main beam is Gaussian with the first sidelobe at around $-25$\,dB. A septum polariser to separate the signal into the two hands of circular polarization has also been designed, built and tested.

\end{abstract}


\section{Introduction}
\label{sec:intro}

The Baryon Acoustic Oscillations (BAOs) from Integrated Neutral Gas Observations (BINGO) telescope \cite{Battye2012,Battye2013,Battye2016,Dickinson2014,Wuensche2018} will observe integrated \hi\ signals  in the redshift range \mbox{$z=0.13$--0.48} to constrain the properties of the Dark Sector \cite{Wang:2016lxa}. It uses the concept of ``single dish, many horns'', with a pair of \mbox{40-m} mirrors and approximately 50 horns in a compact range layout to observe the redshifted 1420\,MHz \hi\ line emission at frequencies of 980--1260\,MHz. Using the \hi\ emission line as a proxy for total matter observation, the BINGO telescope will probe the redshift range where dark energy becomes dominant. This will be complementary to other large-scale structure projects operating at optical or other radio frequencies.

 Very clean optics with low sidelobes is a critical part of the telescope design to achieve the high contrast ratio between the bright foregrounds and the faint \hi\ signals. This contrast will allow component separation techniques to work efficiently \cite{BigotSazy2015,Olivari2016,Olivari2018}, as well as reduce sensitivity to sources of radio frequency interference \cite{Peel2018}. A compact range layout (crossed-Dragone) meets these requirements, but it needs very large horns with apertures of around 2\,m to provide the relatively small beamwidths necessary to appropriately illuminate the secondary mirror. These horns also need to have very low sidelobes to reduce ground pickup.

Manufacturing smooth-sided large horns with narrow bandwidths is relatively simple. Illustrative cases are the horns being used for military radars and the horn used in Bell Labs for the discovery of the Cosmic Microwave Background Radiation \cite{PW1965}. However, a corrugated horn is needed to cover the large frequency range while maintaining the beam performance needed for BINGO. These are common for small to medium-sized horns (e.g., \cite{Witebsky1987,ARCADE.Singal2011}), but not for the \mbox{$>$1\,m-diameter} required for BINGO, leading to the novel construction approach described below. 

An exception is the L-band horn for EVLA \cite{EVLA2005}, produced by the National Radio Astronomy Observatory (NRAO) and briefly discussed at the end of Subsection \ref{subsec:aluminum}. It has a comparable size at similar frequencies to BINGO, but has a different design and uses a slightly different method of fabrication. 

In this paper we summarise the design, construction and testing of a BINGO prototype horn and polariser. Section~\ref{sec:requirements} presents the requirements for the horn with the chosen optical design. Section~\ref{sec:challenges} discusses the challenges of  constructing the large horn and the adopted solution. Sections~\ref{sec:tests} and Section~\ref{sec:polariser} contains, respectively, the methods used for testing the horn and the polariser connected to it, and the results for both tests. In Section~\ref{sec:conclusions} we discuss future work and present our conclusions.

\section{Requirements and optical design}
\label{sec:requirements}

The current optical design for BINGO has a focal length of 63\,m.  In order to slightly under-illuminate the secondary mirror, the horns need to have an aperture of around 2\,m and it is this large size that presents a significant construction challenge. With slight under-illumination of the secondary, the resulting full width at half maximum (FWHM) for the whole telescope is $\sim 40'$, maintaining the original angular resolution proposed in \cite{Battye2013}. The horns need to have very low sidelobes and the beam to be tapered so the illumination is more than $-20$\,dB at the edge of the secondary mirror in order to minimize the spillover, and thus the ground pickup. The resulting very clean beam pattern is vital to enable the faint \hi\ signal to be efficiently separated from the 
bright Galactic foreground emission, whose signal is about 4 orders of magnitude brighter than \hi. Our adopted crossed-Dragone configuration has been a common choice for many Cosmic Microwave Background (CMB) experiments in the last 15 years (see, e.g., \cite{Tran2010,Delabrouille2018,Kashima2018}).

Given the bandwidth of \mbox{980--1260\,MHz}, the design that best satisfies the above requirements across the whole band while having a low ellipticity, a low cross-polarisation and a low return loss, are corrugated horns. As they operate at room temperature, the insertion loss should be less than $-0.1$\,dB across the band, which would limit the increase in the system temperature to around 7\,K. To be well matched with the rest of the system, including the receiver, the horns should ideally have around $30$\,dB return loss across the band.

With such a large aperture setting the beamwidth of the horn, in order to limit its length for manufacturing purposes, we had to choose a profile allowing for a compact horn. Following a previous study \cite{Maffei2004} which led to a design adopted for previous instruments (QUIJOTE and QUBIC for instance, \cite{Hoyland2012,Aumont2016}), a Winston-like corrugated horn design with a circular aperture has been adopted for BINGO. This profile led to a maximum diameter of 1725.9\,mm and length of 4300\,mm (shown in Fig. \ref{fig:profile1}).

The design was optimised using the mode-matching software CORRUG (S.M.T Consultancies Ltd.)\footnote{\url{http://www.smtconsultancies.co.uk/products/corrug/corrug.php}} to model the beam and return loss properties. The corrugations consist of a series of rings with a width (or thickness) of 3.4\,cm, corresponding to 1/8 of the wavelength of the central observing frequency of 1.1\,GHz. The rings have alternating diameters (listed in Table \ref{tab:rings}), creating the corrugation depth (each set for top and bottom of the corrugations). The horn has been designed in order to reach an edge taper below $-20$\,dB at $19^\circ$ across the whole band (horn beam power at the edge of the secondary mirror of the telescope). Together with the low sidelobe level (below $-27$\,dB for the first sidelobe), this results in a spillover below 2\%. 

\begin{figure}[tbp]
   \centering
    \includegraphics[width=0.48\textwidth]{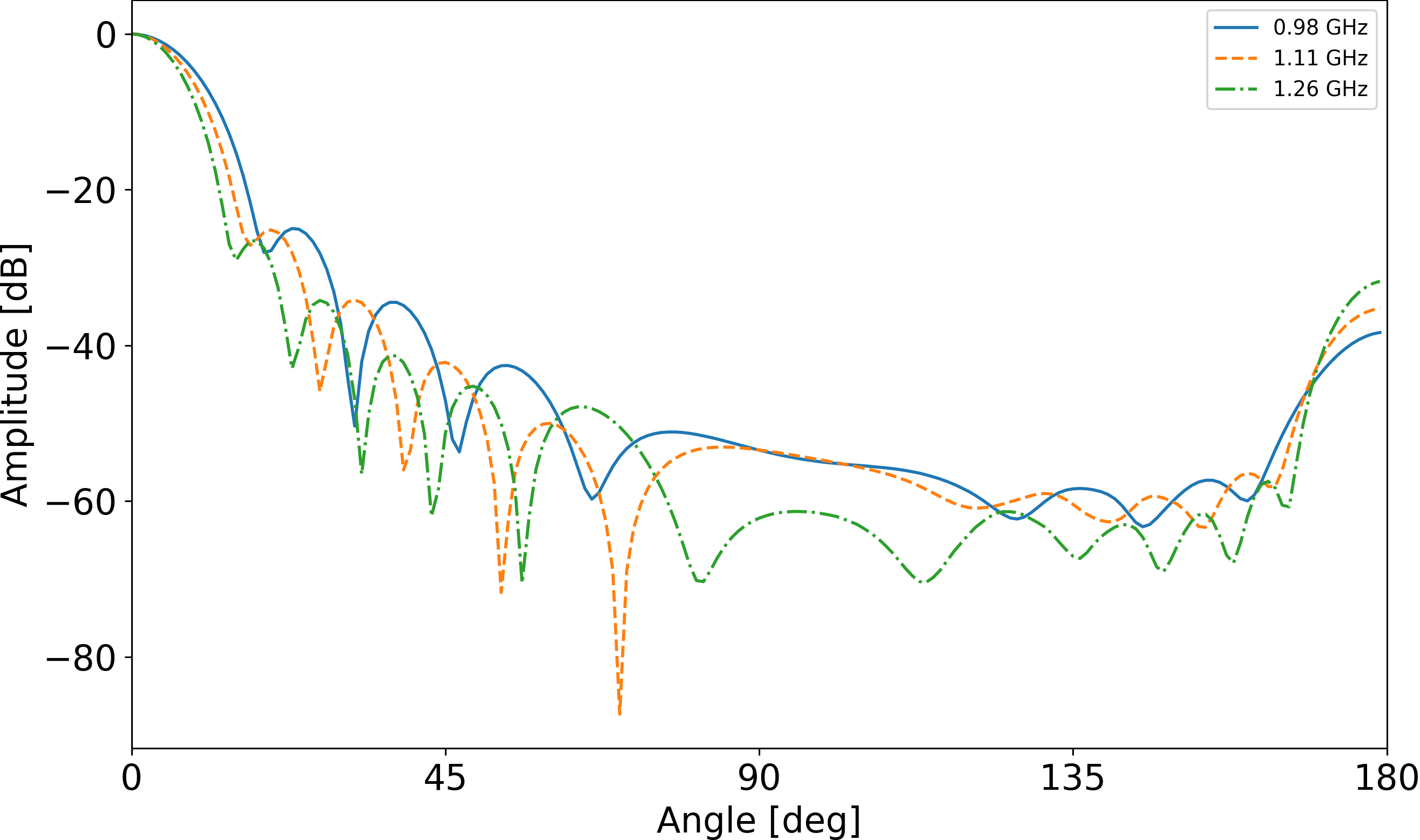}
    \includegraphics[width=0.48\textwidth]{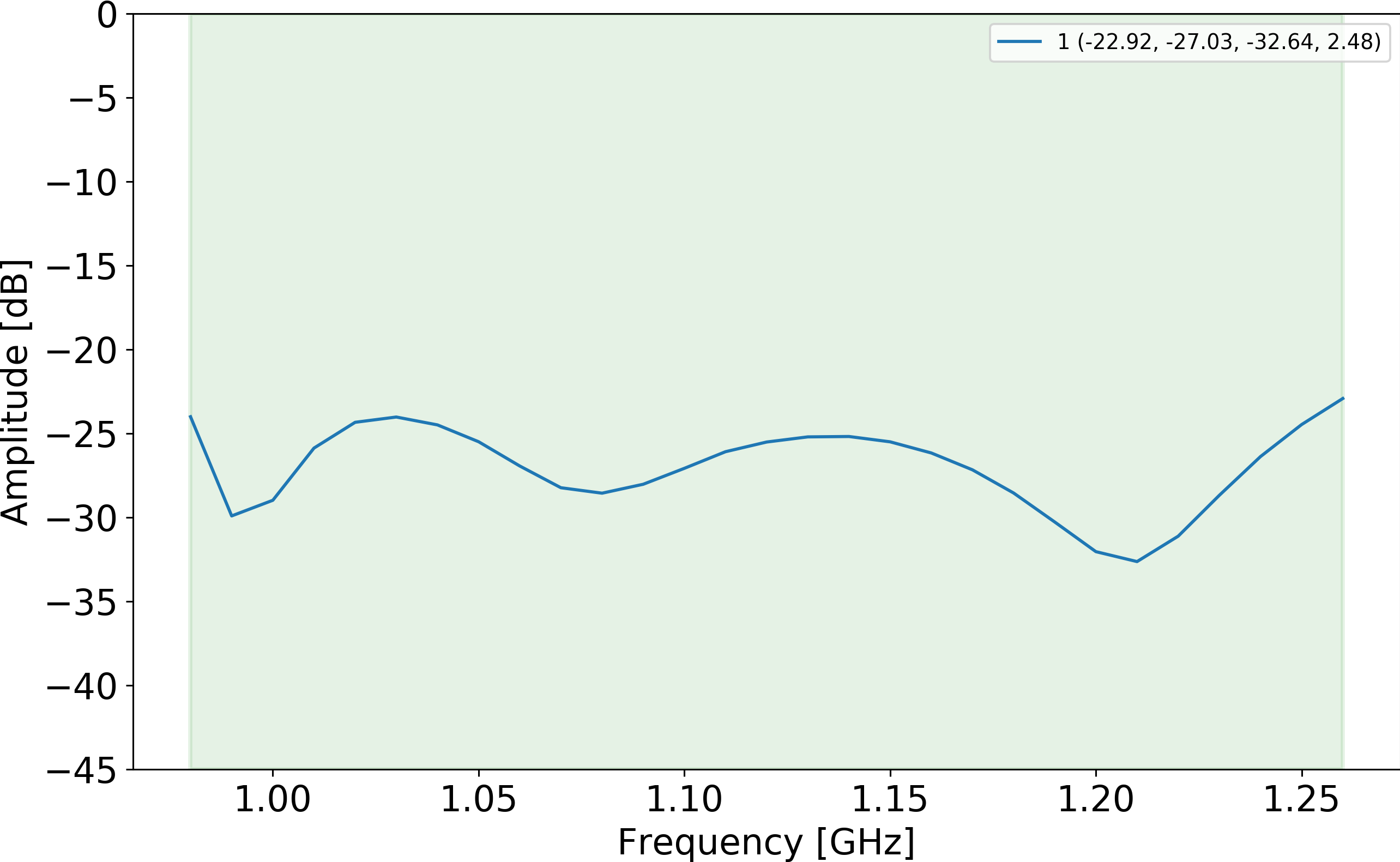}
   \caption{Simulations of the beam pattern for three different frequencies (left) and the return loss (right) of the BINGO horn design} 
   \label{fig:simshorns}
\end{figure}

Simulations of the beam pattern and return loss were then computed using the 3D Computer Simulation Technology (CST) package for a horn with the above ring diameters, and are shown in Figure \ref{fig:simshorns}. The predicted sidelobe performance is excellent with the far-out sidelobes at the $-40$\,dB level. On the other hand, the predicted return loss, particularly at lower frequencies, is acceptable but less than ideal.

The radio emission from the Galaxy and extragalactic sources has a significant fraction of linear polarization. Moreover, the polarization position angle can be frequency-dependent due to Faraday rotation and as a consequence, the measured strength of a linear polarized signal can have an induced frequency dependence that might mimic that from redshifted hydrogen emission. To eliminate this problem it is desirable to make observations using circular polarization. For this reason a polariser is connected to the horn to receive the two hands of circular polarization. The design criteria were that the cross-polarization leakage across the band be less than $-20$\,dB and the return loss be better than $25$\,dB.

\begin{table}[tb]
\caption{The diameters of each ring in the prototype horn}
\label{tab:rings}
\centering
\begin{tabular}{@{}cc|cc|cc|cc@{}}
\hline
\scriptsize
Ring & Diameter & Ring & Diameter & Ring & Diameter & Ring & Diameter \\
(\#) & (mm) & (\#) & (mm) & (\#) & (mm) & (\#) & (mm) \\
\hline
1	&	1725.9	&	33	&	1555.2	&	65	&	1295.7	&	97	&	829.5	\\
2	&	1581\phantom{.0}	&	34	&	1406.2	&	66	&	1138.9	&	98	&	648.3	\\
3	&	1717\phantom{.0}	&	35	&	1542.2	&	67	&	1274.9	&	99	&	784.3	\\
4	&	1571.9	&	36	&	1392.8	&	68	&	1117.3	&	100	&	599.2	\\
5	&	1707.9	&	37	&	1528.8	&	69	&	1253.3	&	101	&	735.2	\\
6	&	1562.6	&	38	&	1379\phantom{.0}	&	70	&	1095\phantom{.0}	&	102	&	545.4	\\
7	&	1698.6	&	39	&	1515\phantom{.0}	&	71	&	1231\phantom{.0}	&	103	&	681.4	\\
8	&	1553.1	&	40	&	1364.9	&	72	&	1071.8	&	104	&	485.7	\\
9	&	1689.1	&	41	&	1500.9	&	73	&	1207.8	&	105	&	621.7	\\
10	&	1543.4	&	42	&	1350.4	&	74	&	1047.7	&	106	&	418.2	\\
11	&	1679.4	&	43	&	1486.4	&	75	&	1183.7	&	107	&	554.2	\\
12	&	1533.4	&	44	&	1335.5	&	76	&	1022.6	&	108	&	339.7	\\
13	&	1669.4	&	45	&	1471.5	&	77	&	1158.6	&	109	&	475.7	\\
14	&	1523.2	&	46	&	1320.1	&	78	&	996.5	&	110	&	244.1	\\
15	&	1659.2	&	47	&	1456.1	&	79	&	1132.5	&	111	&	402.2	\\
16	&	1512.7	&	48	&	1304.4	&	80	&	\phantom{0}969.2	&	112	&	197\phantom{.0}	\\
17	&	1648.7	&	49	&	1440.4	&	81	&	1105.2	&	113	&	382\phantom{.0}	\\
18	&	1502\phantom{.0}	&	50	&	1288.2	&	82	&	\phantom{0}940.7	&	114	&	194\phantom{.0}	\\
19	&	1638\phantom{.0}	&	51	&	1424.1	&	83	&	1076.7	&	115	&	380.6	\\
20	&	1491\phantom{.0}	&	52	&	1271.4	&	84	&	\phantom{0}910.9	&	116	&	197\phantom{.0}	\\
21	&	1627\phantom{.0}	&	53	&	1407.4	&	85	&	1046.9	&	117	&	422.6	\\
22	&	1479.8	&	54	&	1254.2	&	86	&	\phantom{0}879.6	&	118	&	197\phantom{.0}	\\
23	&	1615.8	&	55	&	1390.2	&	87	&	1015.6	&	119	&	386\phantom{.0}	\\
24	&	1468.3	&	56	&	1236.5	&	88	&	\phantom{0}846.6	&	120	&	197\phantom{.0}	\\
25	&	1604.3	&	57	&	1372.5	&	89	&	982.6	&	121	&	410.6	\\
26	&	1456.5	&	58	&	1218.2	&	90	&	811.8	&	122	&	197\phantom{.0}	\\
27	&	1592.5	&	59	&	1354.2	&	91	&	947.8	&	123	&	424\phantom{.0}	\\
28	&	1444.4	&	60	&	1199.3	&	92	&	774.8	&	124	&	197\phantom{.0}	\\
29	&	1580.4	&	61	&	1335.3	&	93	&	910.8	&	125	&	483.2	\\
30	&	1432\phantom{.0}	&	62	&	1179.9	&	94	&	735.6	&	126	&	197\phantom{.0}	\\
31	&	1568\phantom{.0}	&	63	&	1315.9	&	95	&	871.6	&	127	&	466.2	\\
32	&	1419.2	&	64	&	1159.7	&	96	&	693.5	&	128	&	197\phantom{.0}	\\

\hline
\end{tabular}
\end{table}

\section{Challenges of horn fabrication}
\label{sec:challenges}

The traditional method to manufacture corrugated horns for radio astronomy is to use a Computer Numerical Control (CNC) machine to manufacture it from a solid block of material. However, it is very expensive and not practical for the very large horns required for BINGO. Instead, the possibility of constructing them using alternative methods, including layered foam sheets and aluminum profiles bent to produce the corrugations were investigated.

\subsection{Foam sheets}
\label{subsec:foam}

Earlier versions of the BINGO project explored the idea of using sheets of insulating foam. Such sheets are light, inexpensive, widely available and coated with a thin layer of aluminum foil. Holes of the appropriate diameter could be cut into each sheet and the exposed edges covered with copper or aluminum tape to provide a continuous internal conducting surface. Initial tests with a horn of diameter 0.55\,m made up of 78 of these sheets gave excellent results, with insertion loss around $-0.1$\,dB. Unfortunately scaling up to a diameter of around 1\,m was not so successful: similar performance levels could be achieved for short periods but not reliably maintained from day to day. Additionally, such sheets are likely not the best choice to perform under the high temperatures (year average of $26^{\circ}$C) on site during the full mission. The findings above led to the search of an alternative approach.

\subsection{Aluminum profiles}
\label{subsec:aluminum}
\begin{figure}[tbp]
   \centering
    \includegraphics[width=1.0\textwidth]{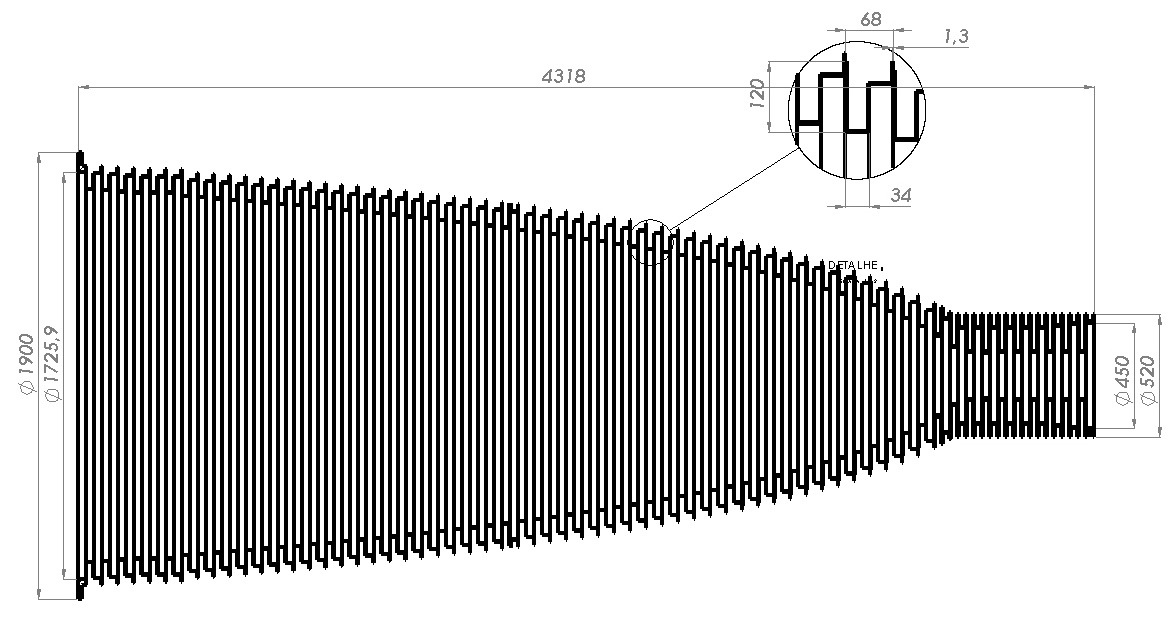}
   \caption{Engineering drawing for the horn. The insert contains the ``upside-down chair'' profile used to produce the desired corrugation. Dimensions are in mm.}
   \label{fig:profile1}
\end{figure}

\begin{figure}[tbp]
   \centering
   \includegraphics[width=0.48\textwidth]{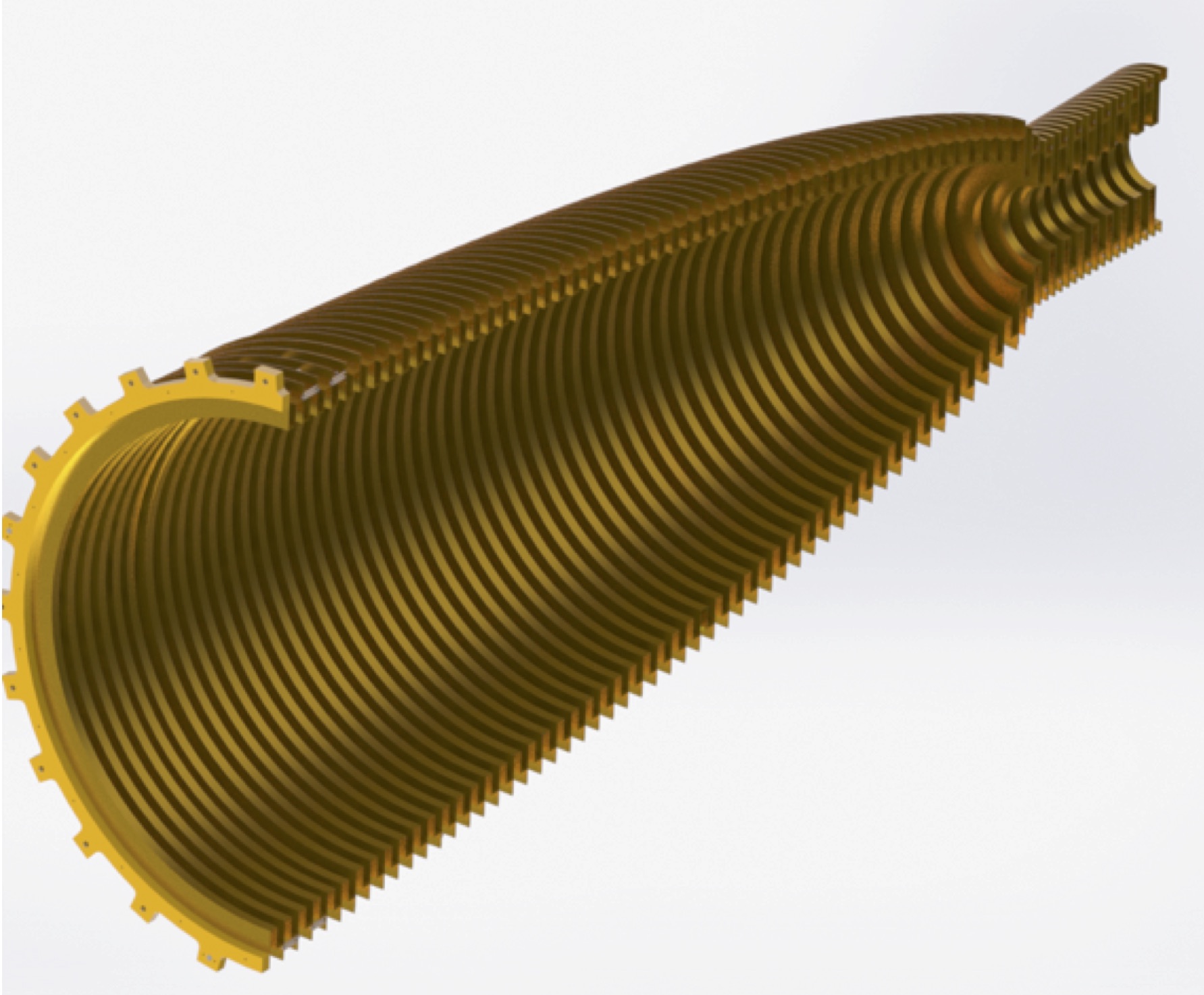}
   \includegraphics[width=0.48\textwidth]{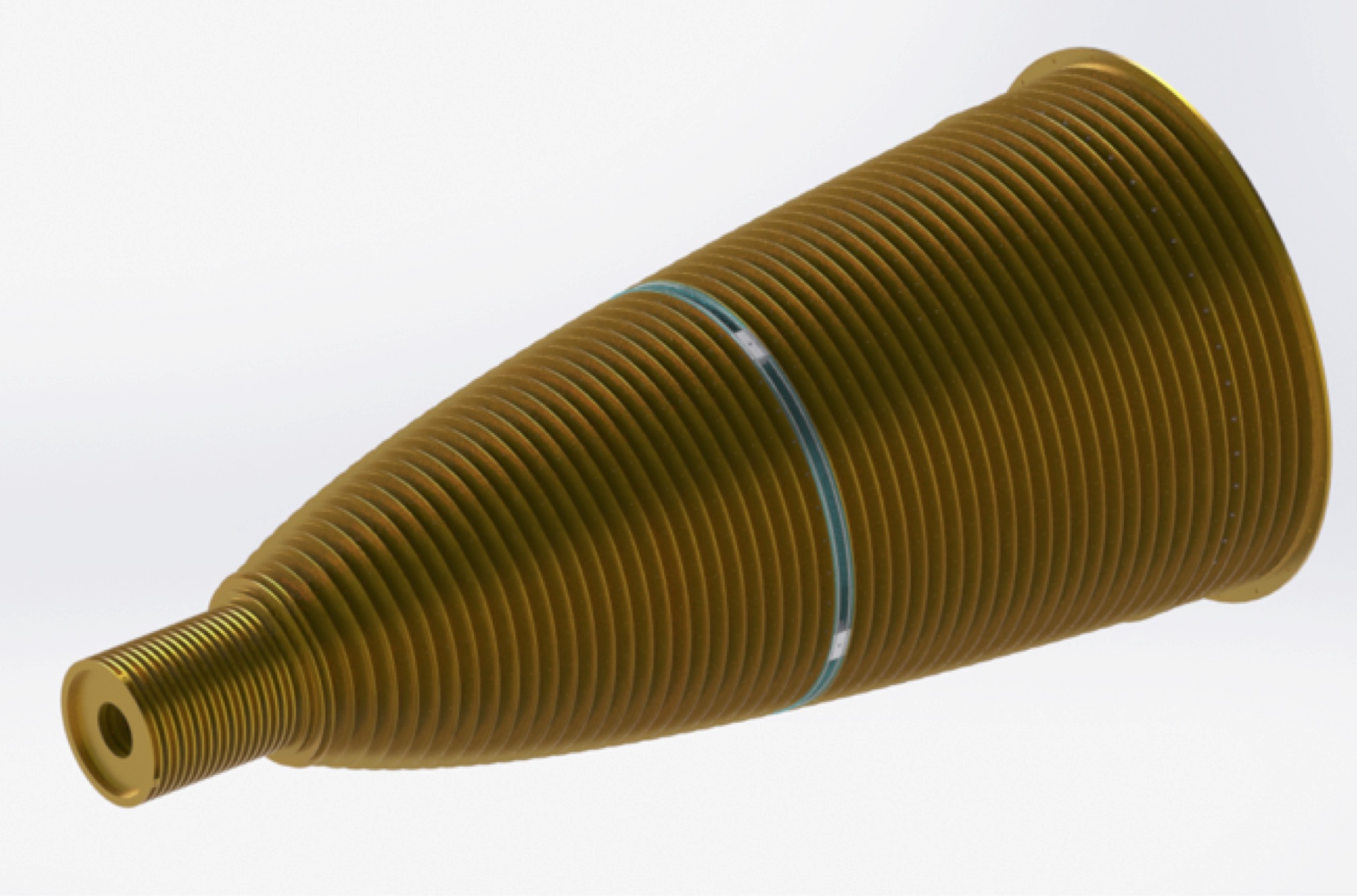}
   \caption{Left: horn longitudinal cut through ring diameters. Right: Opposite view, with a suspension point in a machined ring in the middle of the horn (green ring).}
   \label{fig:horndesign}
\end{figure}

One of the alternative construction approaches investigated, and which was considered the most promising in terms of costs and delivering a lightweight unit, is to bend relatively thin chair-shaped sections (see inset in Figure \ref{fig:profile1}) of aluminum into circles of appropriate diameter and then join them together. This technique was used to produce all the larger diameter rings for a corrugated horn prototype, except for the largest one, which was manufactured using CNC machines for rigidity. The throat section of the horn was made using annuli cut from aluminum sheets, with a sheet of aluminum bent around to form the centre of the horn ring, welded together to make a ring. The centre mounting ring, and the last throat ring, were made with a CNC machine, again for rigidity (see Figure \ref{fig:horndesign}). The rings were joined together either by welding, riveting or bolting depending on ring diameter. 

Preliminary studies indicated that two families of aluminum alloys (5000 and 6000) might have the required properties to be formed into the shaped profile. A high level of malleability of the aluminum alloy used for fabrication was required to withstand cold mechanical conformation without weakening the material, particularly for the smaller radius profiles. Also relevant is the hardness of the material, the thickness and internal radii of the profile. This is important in order to allow the bending of the rings without causing undesirable deformation stress at those edges and the ease at which it can be welded to close the rings while minimising the deformations in the bent profile. The chosen alloy was Al5052 and, based on the properties of this material, an appropriate tool to extrude the profiles could be designed.

\begin{figure}[tbp]
   \centering
   \includegraphics[width=0.32\textwidth]{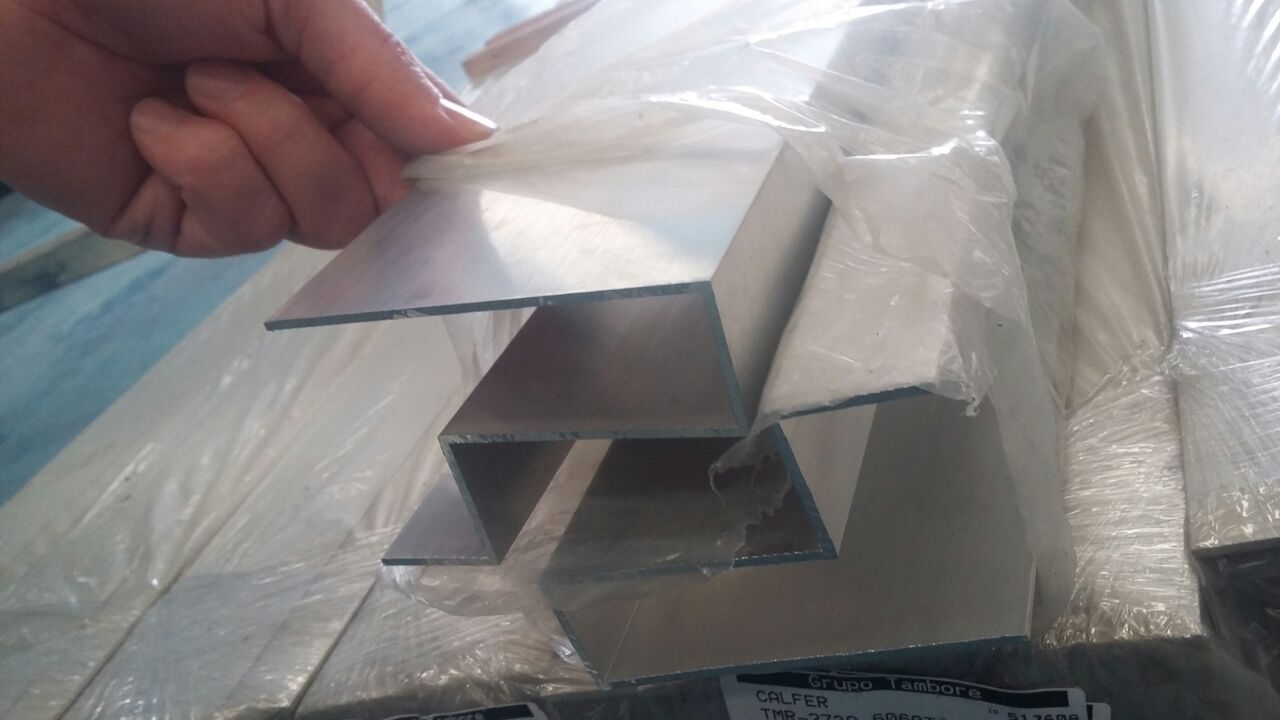}
   \includegraphics[width=0.32\textwidth]{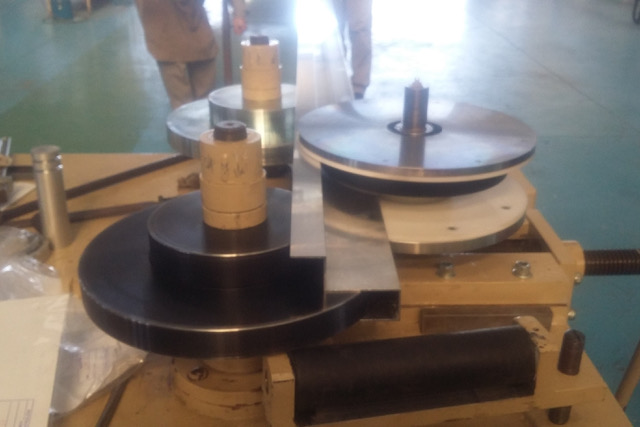}
   \includegraphics[width=0.32\textwidth]{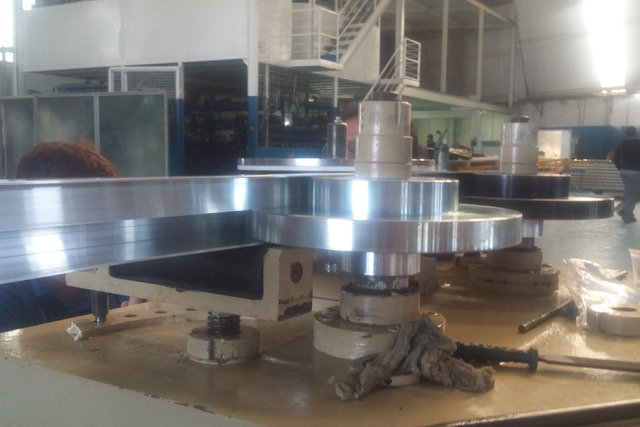}
   \includegraphics[width=0.32\textwidth]{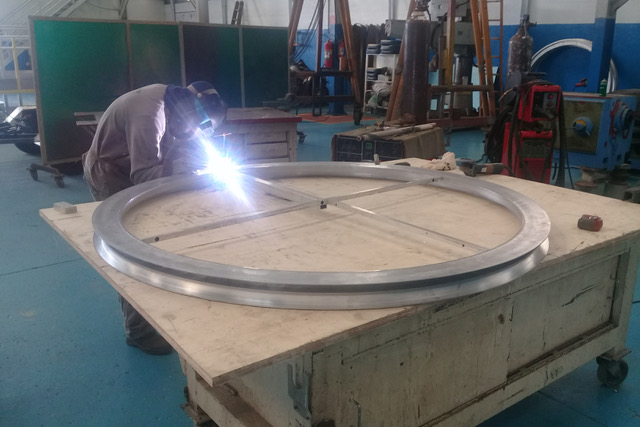}
   \includegraphics[width=0.32\textwidth]{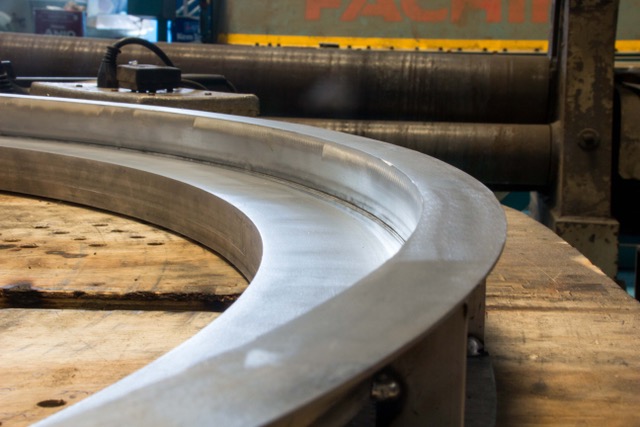}
   \includegraphics[width=0.32\textwidth]{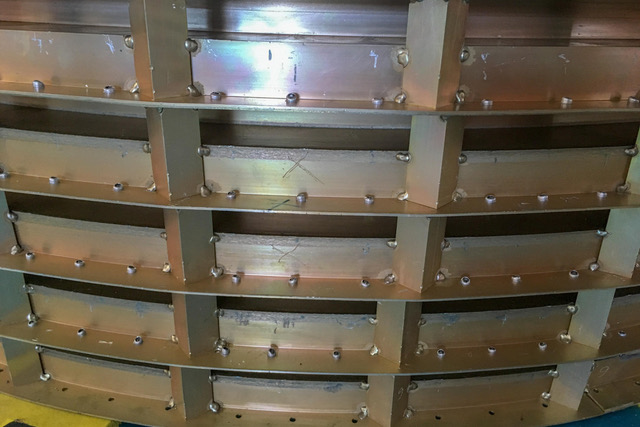}
   \caption{The horn construction process. Starting from chair-shaped profiles bent out of a single sheet of aluminum (top-left), these profiles are then bent into rings. Top-middle: View of the 3 rollers used for conformation. Top-right: Side view of the fitting of the profile onto the rollers. Before being welded together to produce a complete ring (bottom-left) and any deviations smoothed out (bottom-middle). They are then stacked and riveted to each other (bottom-right) to form the complete horn}
   \label{fig:horns1}
\end{figure}

Calfer\footnote{\url{http://calfer.com.br/}}, a spin-off company from the Brazilian airplane manufacturing industry, was contacted to produce a prototype horn. This included the manufacturing of the proper tool to bend the profiles into shape according to the required dimensions and tolerances. The profiles are shown in Figure \ref{fig:horns1} (top-left). The chair-shaped material with these profiles was then fed through a bending machine using rollers that were specially designed to accommodate the profile (Figure \ref{fig:horns1}, top-middle and top-right). Due to the size and tolerances (better than 1\,mm deviation from designed ring dimensions) required by the BINGO design, it is essential that small deformations, usual in the bending process, are minimized at the bending and welding stages, reducing finishing time and final costs.

A single bending tool with three wheels can be adjusted to make rings with a range of diameters. The end of the rings are welded together, making use of a rigid template for proper positioning (Figure \ref{fig:horns1}, bottom-left). Heating of the material during the welding slightly deforms the nearby areas, thus requiring some rework before assembling the rings (Figure \ref{fig:horns1}, bottom-middle and bottom-right).

Before assembling the parts, there must be a careful dimensional and visual inspection of each ring, since access to the interior of the horn is difficult after the completion of the assembly process. The inner surface finish of each ring plays an important role in the final performance of the horn. The rings should not have significant irregularities on their surfaces and were sanded to a smooth finish. The dimensions were checked on a gauge table after the welding process.

The final assembly of the rings to form the horn was done manually. The best alignment of the center of all rings was achieved after careful calibration of the diameters. This ring alignment operation is extremely important because it can directly influence the electromagnetic performance. Two of the rings, one at the aperture of the horn and one in the middle,  were reinforced to act as a rigid structure to receive lift lugs for maneuvering the horn after assembly. The prototype manufactured in this way has a total mass of about 448\,kg (including bolts and lift lugs) with a build accuracy equal to or better than 1.0\,mm on average. 

As a comparison, the NRAO E-VLA L-band horn is slightly smaller than the BINGO horn (4.1\,m length; 1.5\,m aperture diameter). Their horns were made by stacked aluminum rings and bands, made of separate aluminum alloys, held together with a fiberglass shell on the outside \cite{EVLA2002-a}, \cite{EVLA2002-b}. One of the aluminum alloys used by NRAO is the same as the one adopted in this work (6061 and 5052 for the NRAO E-VLA versus 5052 only, for the BINGO rings). Their assembly process is described in \cite{EVLA2005}, and consists of setting up pre-produced corrugations in an assembly fixture provided by NRAO. These corrugations are a mix of fully machined rings and rings made by a combination of sheet metal disks and bands, which also work as flanges to assemble different sections of the horn. 


\section{Horn tests}
\label{sec:tests}

There are three main radio frequency test procedures required to characterise the performance of a horn. The first is to measure the insertion loss, i.e., the loss of the signal as it passes through the device. The second is to measure the return loss, i.e., how much signal is reflected by the horn due to small impedance mismatches. These measurements of loss were made using an Agilent N5230C PNA-L desktop Vector Network Analyser (VNA). The relevant S-parameter is S11; when a horn is terminated by a metal reflector S11 gives the insertion loss and when terminated by an RF absorber, in our measurements a material with trade name Eccosorb, S11 measures the return loss.The third measurement on the horn is the characterisation of the beam pattern which is done using a specialised feed testing facility. These need to be measured across the frequency range over which the horn is required to operate. Also in this section we describe measurements of the waveguide transitions which were required to make the horn polar diagram measurements.  All the loss measurements for transitions were made with the same VNA. 



\subsection{Transitions}
\begin{figure}[tbp]
   \centering

\includegraphics[width=0.48\textwidth]{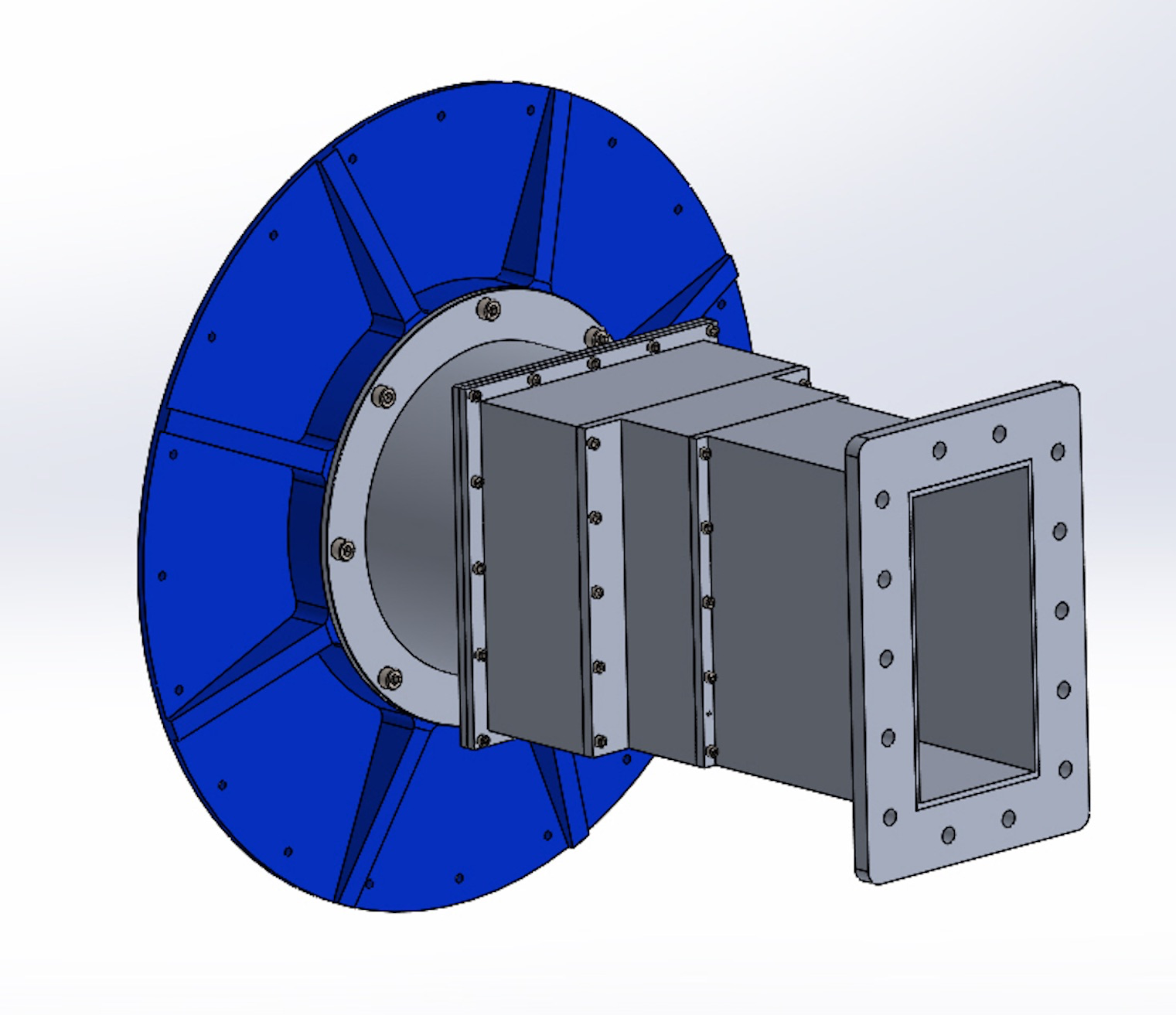}
\includegraphics[width=0.48\textwidth]{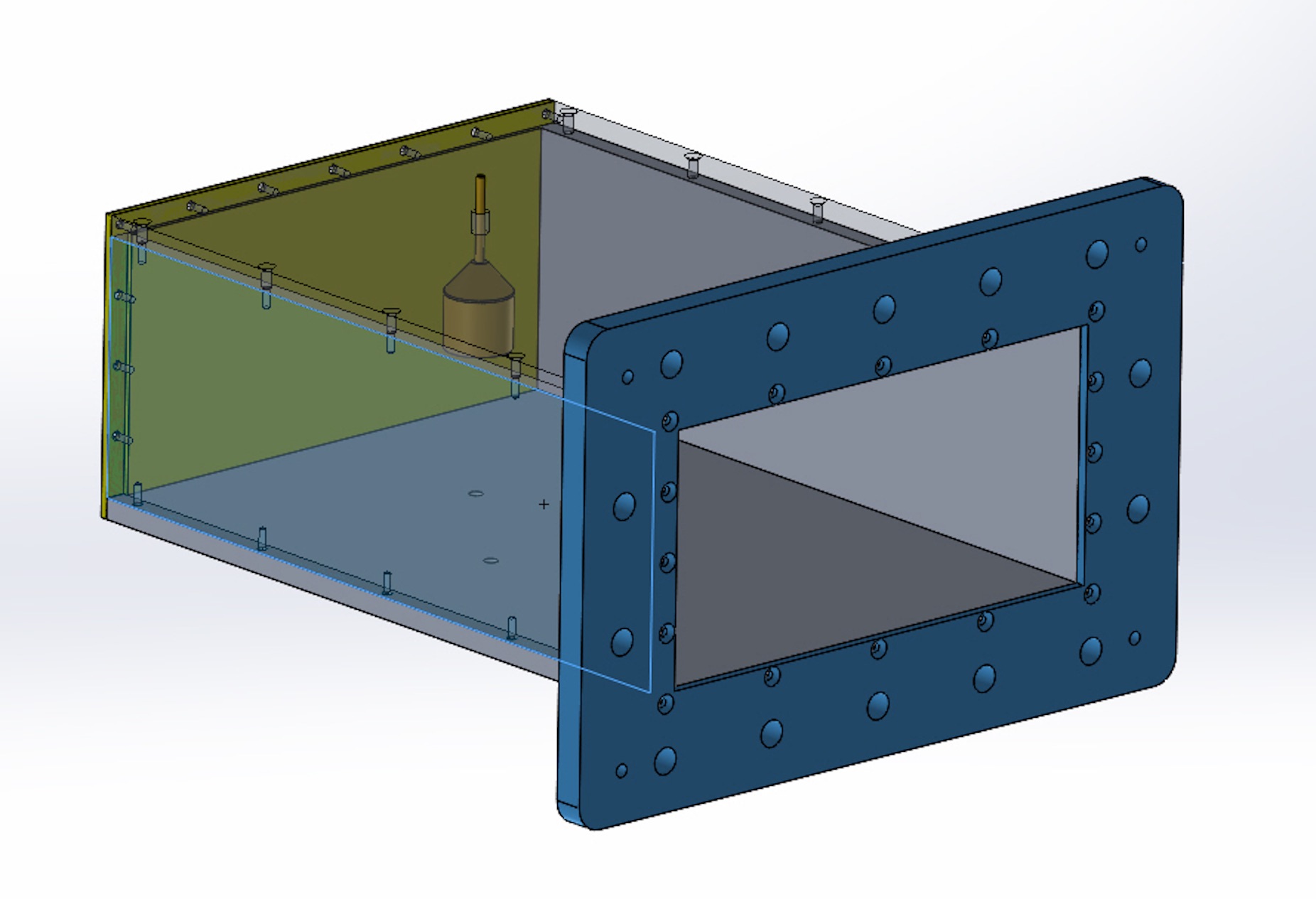}

\caption{Left: Circular to rectangular transition used for polar diagram measurement. Right: Rectangular (WG5) to coax transition, showing the internal antenna. The bobble has an internal thread that allows for a precise adjustment inside the cavity. The WG5 transitions were used both for polar diagram and magic tee measurements.}
\label{fig:circ2wg52coax}
\end{figure}


A selection of waveguide transitions is required for different purposes. A circular-to-rectangular transition was required to connect the horn to a
rectangular to coax (WG5) transition, allowing for the interface with the test equipment.  The circular to rectangular transition was manufactured at Calfer and the WG5 transition, at Phase2 Microwave. The rectangular-to-coax transition has internal dimensions of $97.79 \times 195.58$\,mm and the circular-to-rectangular transition consists of three sections. The WG5 to coax drawings and the circular-to-rectangular transition can be seen in Figure \ref{fig:circ2wg52coax}

Input and return loss measurements can easily be affected by errors in the VNA calibration. In particular, any losses in the connecting cables and waveguide transitions connected to the horns need to be measured and calibrated out to give just the horn properties. Initial measurements used the standard VNA calibration components at the end of the SMA cables, and the loss of the waveguide components (except for the horn) has to be subtracted from the measurements later on. The insertion loss of the WG5 transition was measured with the VNA, and we obtained losses between $-0.025$ and $-0.12$\,dB (mean of $-0.075$\,dB) in the BINGO band. The measured rectangular-to-circular insertion loss figures were between $-0.06$ and $-0.15$\,dB (mean of $-0.1$\,dB). Once this was done the actual horn return and insertion loss were measured. 

However, the standard cable calibration procedure was not accurate enough for the tests, so we looked for a new procedure, aiming for higher accuracy. A waveguide calibration, including the WG5 transitions in the calibration process, was tested using a 1/8 wavelength waveguide offset and a rectangular blanking plate (``short'') manufactured specifically for this new process. This method was preferred for all future measurements of the waveguide components.

\subsection{Horn prototype}
\begin{figure}[tbp]
   \centering
\includegraphics[width=0.32\textwidth]{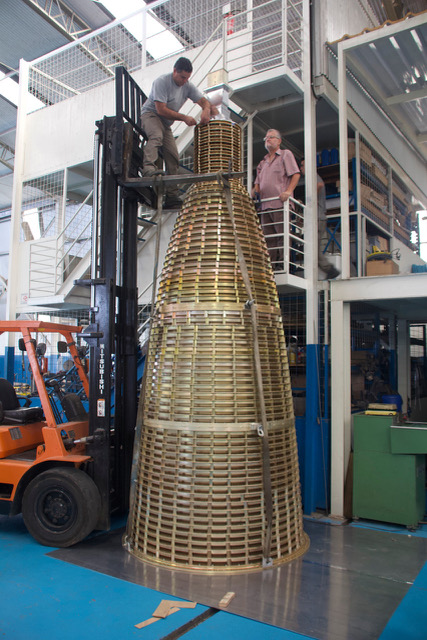}
\includegraphics[width=0.32\textwidth]{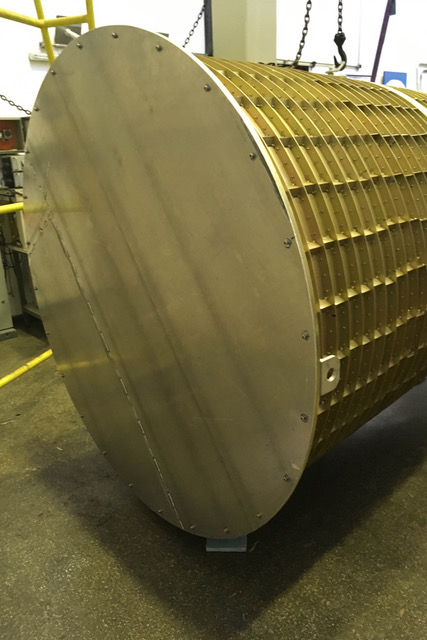}
\includegraphics[width=0.32\textwidth]{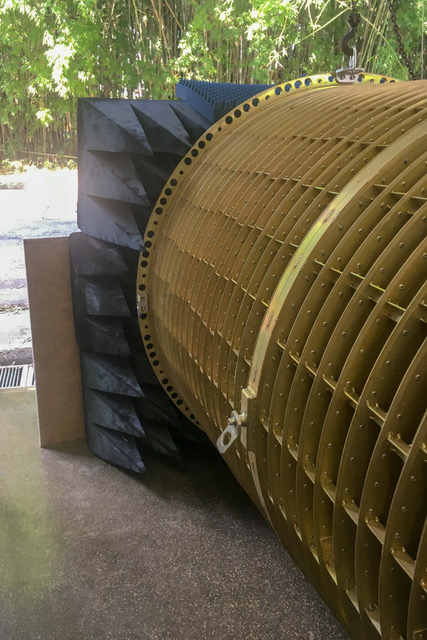}
\caption{Left: Initial horn testing at Calfer. Middle: Testing the horn at INPE with a aluminum plate covering the horn aperture. Right: At INPE with Eccosorb covering the horn aperture}
\label{fig:horns_mouthcovered}
\end{figure}

Initial measurements of the insertion loss of the horn were made at Calfer (Figure \ref{fig:horns_mouthcovered}, left panel) to check that the horn was functioning correctly before acceptance. These were followed by more intensive tests during 2018 of both the insertion and return loss at INPE, using cable calibration. The final set of measurements presented here were taken in February and March 2019 and used waveguide calibration.

\begin{figure}[tbp]
   \centering
   \includegraphics[width=0.48\textwidth]{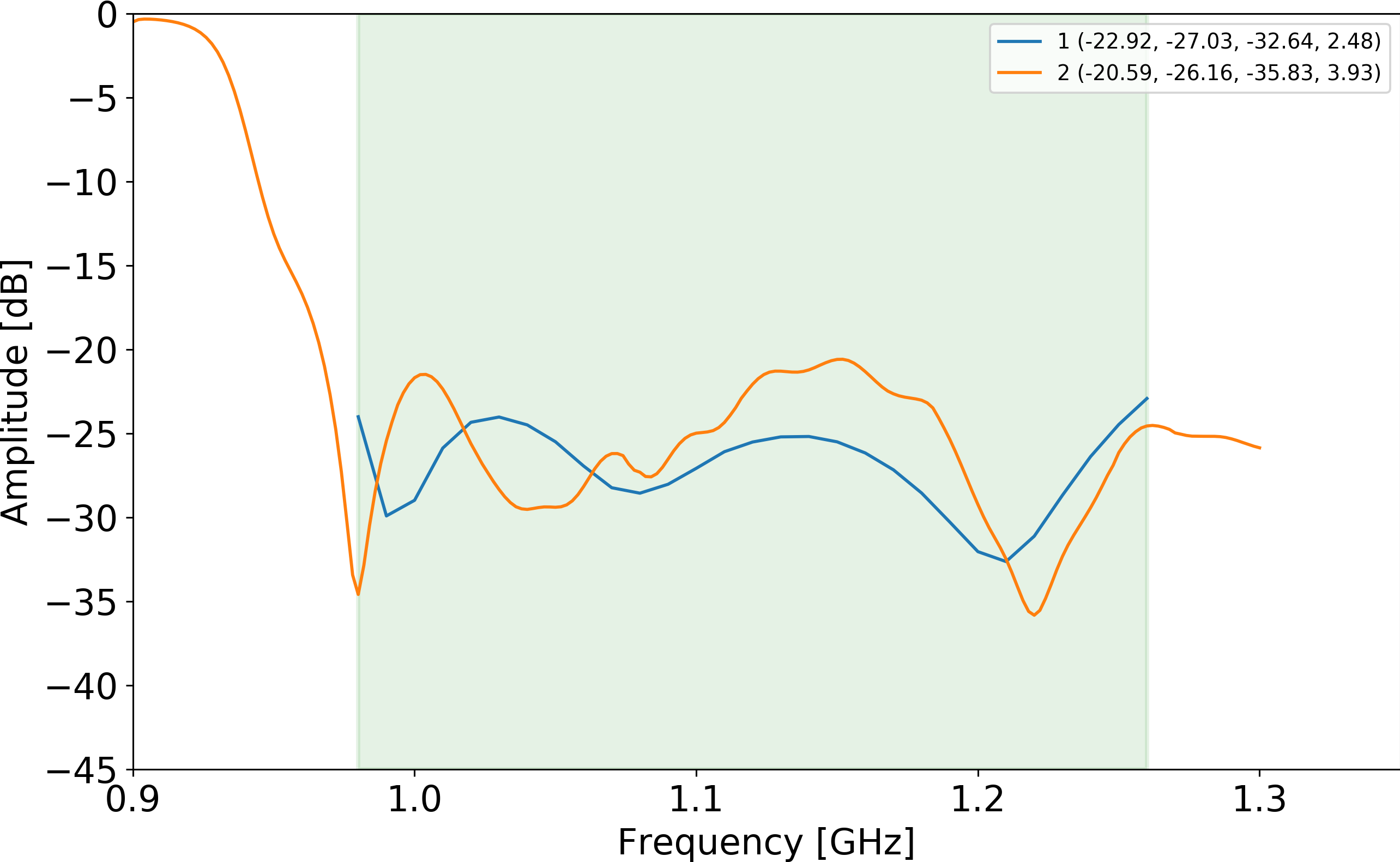}
   \includegraphics[width=0.48\textwidth]{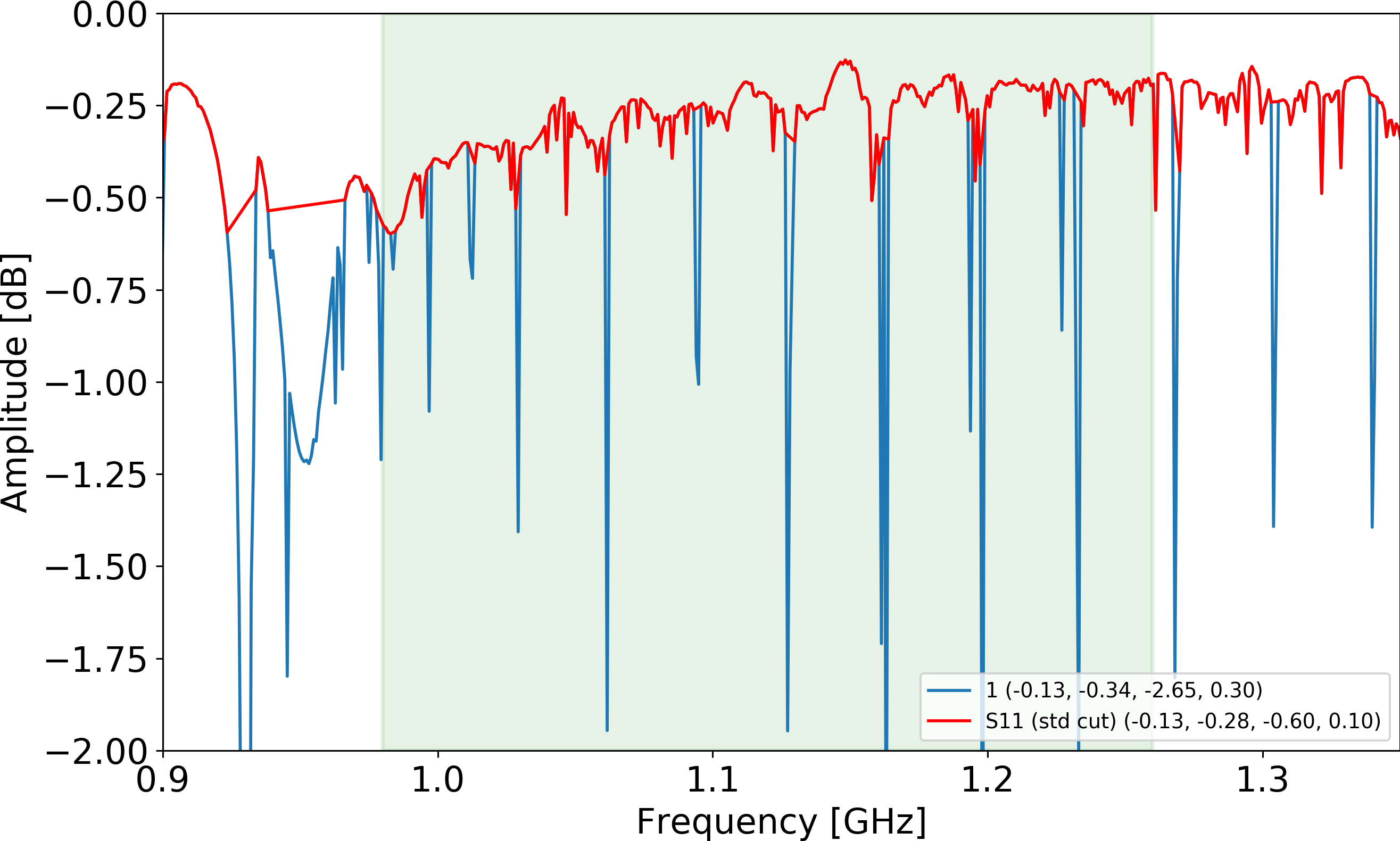}
   \caption{Left: The return loss of the prototype horn. Right: The insertion loss of the prototype horn, with waveguide calibration. The measurement is of both the circular-to-rectangular transition and the horn, with the reflector at the end, so the two passes double the values. The raw measurement is shown in green, after spike removal is shown in red. The numbers in the legend are the maximum, mean, minimum and standard deviation within the BINGO band of 980--1260MHz (shown by the green shaded region)}
   \label{fig:final_loss}
\end{figure}

The final insertion loss measurements of the horn including the circular-to-rectangular transition are shown in Figure \ref{fig:final_loss}. It will be noted that there are regular spikes in the insertion loss, which arise from standing waves in the formed ``cavity'' between the reflecting plate and some part of the horn throat. These spikes will not be present in normal operation without the reflecting plate and can be ignored. If we do this the measured insertion loss within the BINGO band is $-0.14$\,dB, with a standard deviation of 0.05\,dB. We estimate the uncertainty on this measurement due to VNA calibration to be around 0.05\,dB. This loss would result in an increase in system temperature of around 10\,K. Subsequently, we have identified areas in which small modifications in the fabrication process can be made to try and reduce this loss, and these will be tested with a second prototype.

\subsection{Beam measurements}
\label{LIT}

\begin{figure}[tbp]
   \centering
   \includegraphics[width=0.4\textwidth]{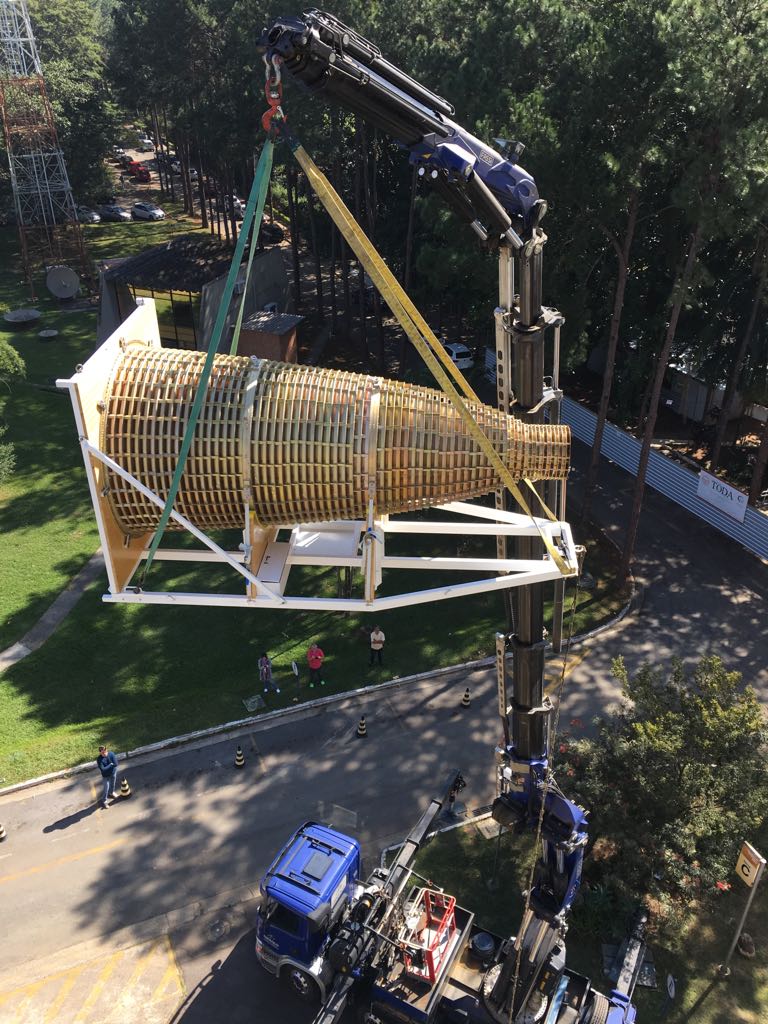}
   \includegraphics[width=0.4\textwidth]{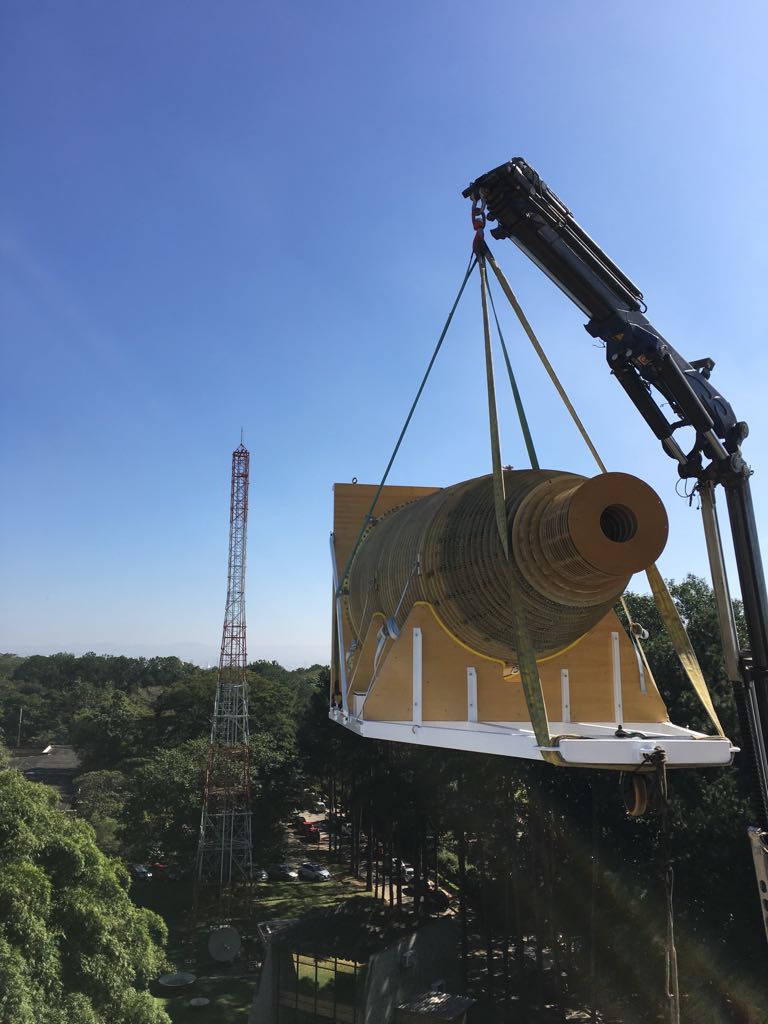}
   \includegraphics[width=0.4\textwidth]{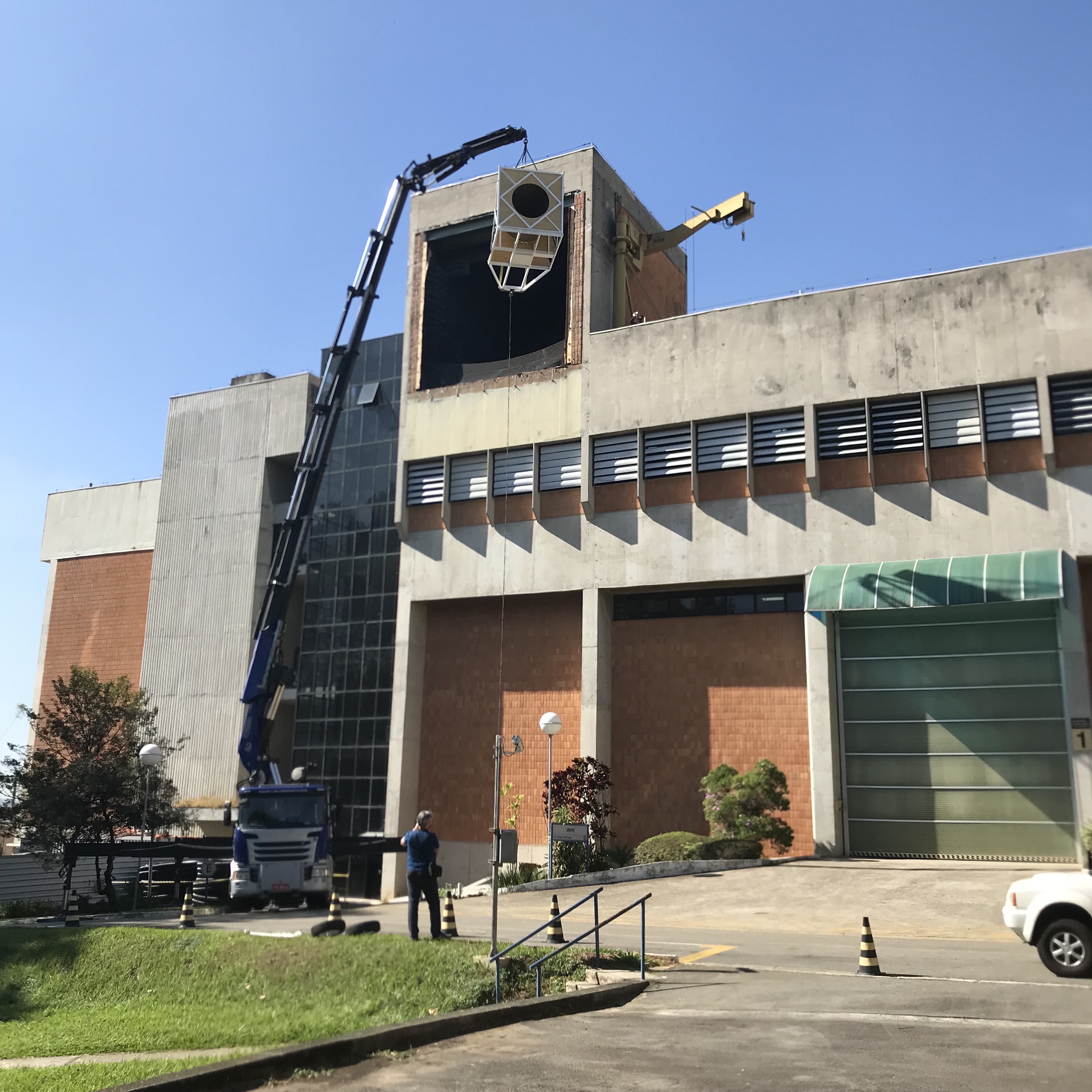}
   \includegraphics[width=0.4\textwidth]{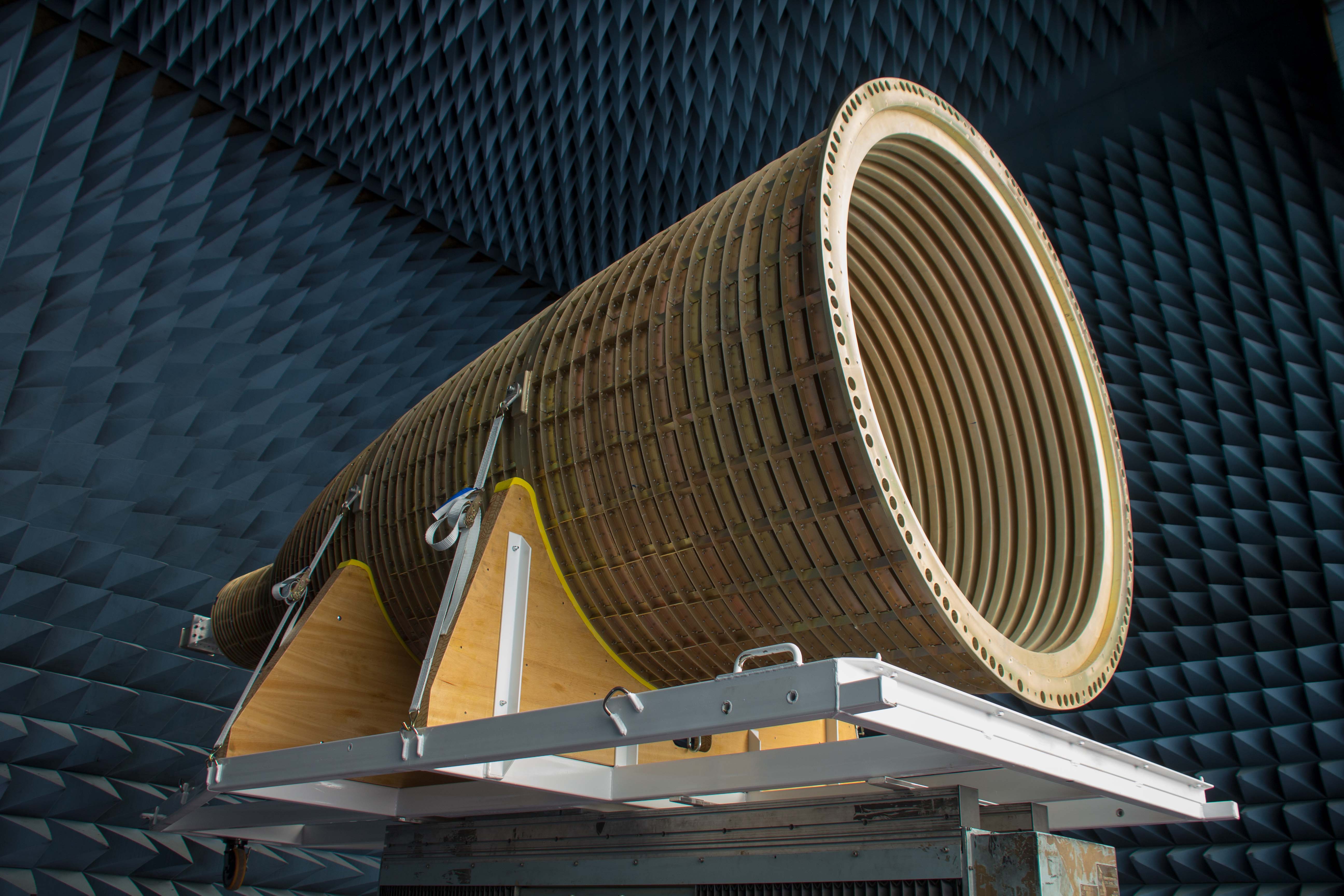}
   \caption{Top left, top right and bottom left: Horn being suspended, with its cart, to the opening of the anechoic chamber of LIT test facility. Bottom right: The horn mounted inside the anechoic chamber.}
   \label{fig:horn_lit1}
\end{figure}

\begin{figure}[tbp]
   \centering
   \includegraphics[width=0.48\textwidth]{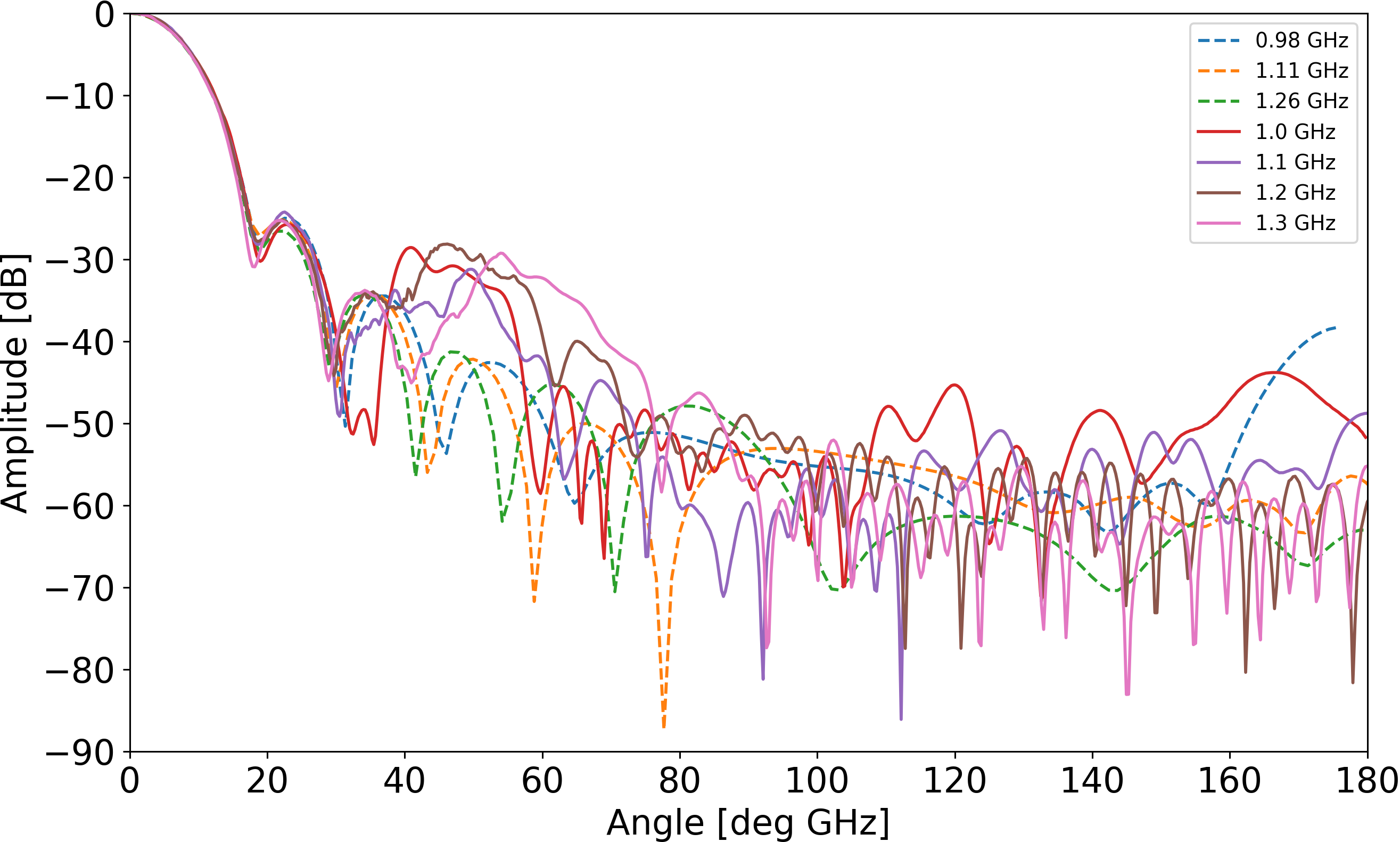}
   \includegraphics[width=0.48\textwidth]{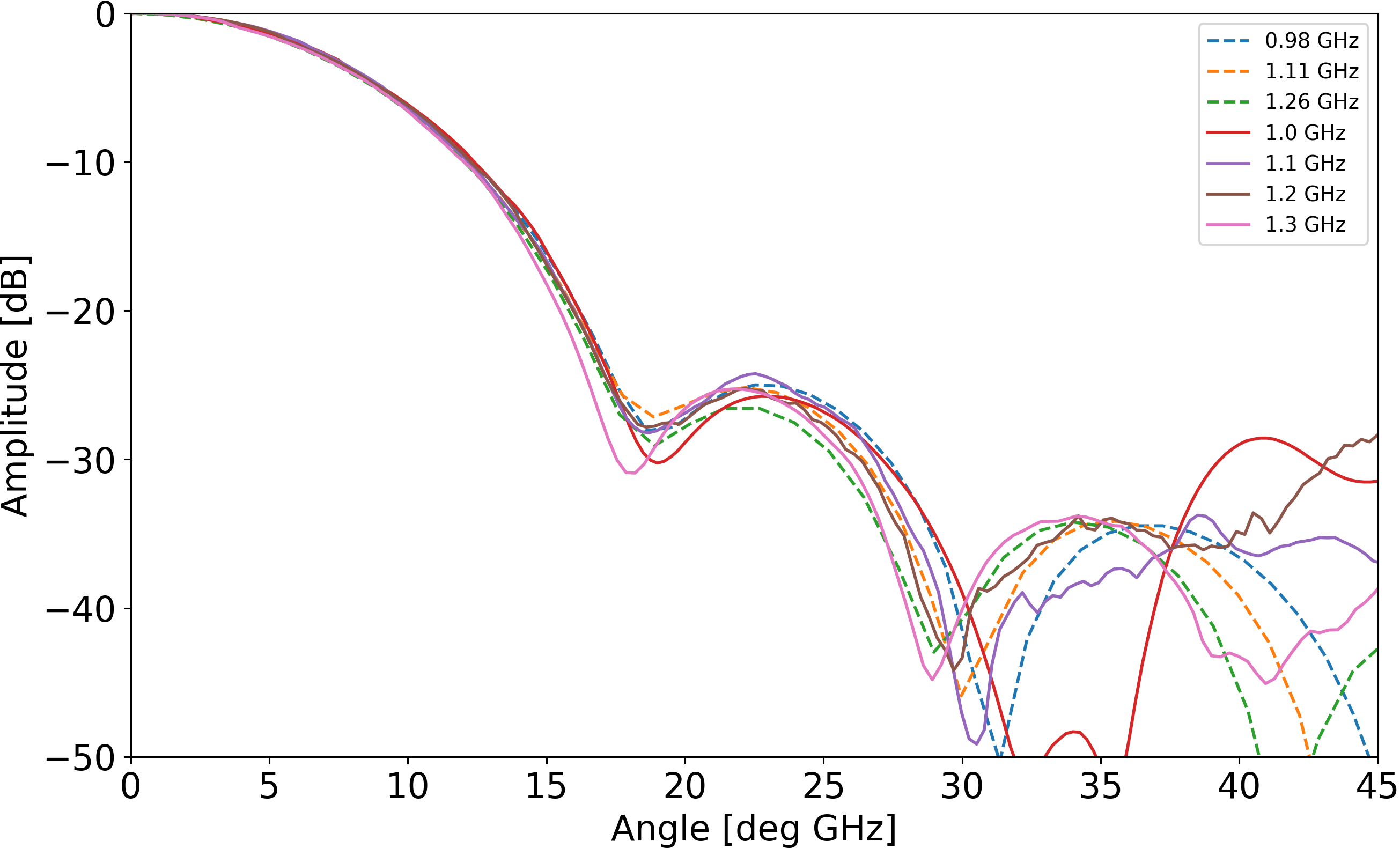}
   \caption{The beam patterns at a range of frequencies across the BINGO band, rescaled in width to 1\,GHz and in amplitude to a maximum of 0\,dB. Left: full azimuth range. Right: a zoom-in on the first two sidelobes. The first sidelobe is around 25\,dB lower than the main beam
   }
   \label{fig:horn_lit2}
\end{figure}

\addtolength{\tabcolsep}{-4pt}
\begin{table}[tb]
\caption{Beam measurements. Left: horizontal polarization. Right: vertical polarization. Minimum rejection was not measured at all frequencies}
\scriptsize
\centering
\begin{tabular}{@{}c|cccc|cccc@{}}
\hline
Freq 		& Directivity 	& FWHM 	& Min. 	& Front to back	& Directivity 	& FWHM 	& Min. 	& Front to back \\
	 	& 	 		& 	 	& rejection  	& difference	& 	 		& 	 	& rejection  	& difference	\\
(MHz)	& (dB)		& (deg)	& (dB)		& (dB)		&  (dB)		& (deg)	& (dB)		& (dB)		\\	
\hline
\phantom{0}950 & 22.4 & 14.6 & -- & 47.0 & 22.7 & 13.4 & -- & 56.2\\
1000 & 23.1 & 13.8 & 22.5 & 51.7 & 23.0 & 13.5 & 20.5 & 64.7\\
1050 & 23.5 & 13.1 & 28.2 & 54.2 & 23.5 & 12.9 & 35.9 & 57.2\\
1100 & 23.9 & 12.4 & 25.6 & 58.2 & 23.9 & 12.4 & 30.5 & 56.2\\
1150 & 24.3 & 11.8 & 20.0 & 72.8 & 24.4 & 11.7 & 19.7 & 55.4\\
1200 & 24.7 & 11.1 & 26.4 & 57.2 & 24.6 & 11.1 & 26.0 & 63.6\\
1250 & 25.2 & 10.3 & -- & 56.1 & 24.9 & 10.6 & -- & 57.1\\
\hline
\end{tabular}
\label{tab:beam_measurements}
\end{table}
\addtolength{\tabcolsep}{4pt}

The Laboratory of Integration and Tests (LIT) is a Brazilian facility to integrate and test satellites up to the 2-ton class, located at INPE, in S\~ao Jos\'e dos Campos, Brazil. It has a far-field antenna test range that was used to measure the horn beam pattern. Measurements of the prototype BINGO horn were conducted during 22--25 May 2018. Figure \ref{fig:horn_lit1} depicts the procedure of inserting the prototype horn inside the antenna test range and its final positioning facing the transmitter before the beginning of the tests.

The test facility consists of a 9\,m wide $\times$ 9\,m tall $\times$ 10\,m deep chamber, fully covered inside with absorbing cones and facing the transmitting antenna through a 9\,m $\times$ 9\,m open window. The transmitter is mounted on a tower located 80\,m away from the chamber and can produce both horizontal and vertical linearly polarized signals, rotating the transmitting feed by $90^{\circ}$. The horn was mounted on top of a turntable with its aperture located about 1.5\,m ahead of the rotation axis. This system is capable of continuous $360^{\circ}$ horizontal rotation, clockwise and counterclockwise, and the horn response to the signal emitted by the far-field transmitted is continuously measured until a $360^{\circ}$ horizontal rotation is completed. H polarization measurements were taken with the transition attached to the throat and its longest length in the vertical position. V polarization measurements were taked with the transition rotated by $90^{\circ}$ in respect to the former position. Chamber temperature and humidity were not controlled nor taken into account during the data acquisition process, but should not have a significant effect on these measurements.

The radiation pattern measurements were performed using linear polarization transmitters on the tower; the signal was generated by a ANRITSU MG3649B equipment. The horn was connected to a MI-1797 receiver (fabricated by Scientific Devices Australia Pty. Ltd. - SDA) via the same circular-to-rectangular and rectangular-to-coax (WG5) transitions as described in the previous subsection. The acquisition software is also from SDA. We made 17 measurements at 25\,MHz intervals from 900 to 1300\,MHz (900, 925, 950, ..., 1250, 1275, 1300 MHz), and each frequency was measured in both vertical and horizontal direct polarization, with simultaneous cross-polarization measurements in 7 of them (900, 1000, 1050, 1100, 1150, 1200, 1300 MHz). The directivity was calculated by numerically integrating the irradiation diagrams.

Table \ref{tab:beam_measurements} shows a sample of the data for the frequencies where E-plane (direct) and H-plane (cross) measurements were made simultaneously. Directivity and FWHM are essentially the same for both polarizations in each frequency, indicating that the beams should be nearly circular across the frequency band. 
The front to back relation refers to the ratio between signal measured at $0^{\circ}$ and signal measured  at $\pm 180^{\circ}$ and indicates the level of rejection of the horn beam pattern. There is a spread in the values for vertical (about 15\, dB) and horizontal polarization (25 \,dB). Nevertheless, the worst case (for horizontal polarization at 900\,MHz) is 46\,dB, and, inside the nominal band (980--1260\,MHz), the worst response is 48.39\,dB, also for horizontal polarization. 

Beam patterns are shown in Figure \ref{fig:horn_lit2}, with the performance along a full azimuth scan in the left-hand panel, and a zoom over 0--45$^{\circ}$ in the right-hand panel. The beamsize of the horn alone depends on frequency, while the reduction in illuminated mirror surface will compensate for this to first order so that the telescope beam on the sky will be relatively frequency-independent. As such, we scale the beam width by the frequency in GHz to normalise the beam width to 1\,GHz.

Minimum rejection refers to the cross-polarization (leakage from V to H during H measurements and vice-versa) contamination and was measured at 900, 1000, 1050, 1100, 1150, 1200 and 1300\,MHz. Transmitter in the antenna was rotated by $90^{\circ}$ to verify cross-polarization sensitivity. The worst case is a minimum rejection of 19.7 dB at 1150\,MHz in the vertical polarization.

Simulations of the cross-polarization performance for this specific design indeed shows a much better performance for cross-pol rejection. However, we have no reason to believe in a poor horn design, due to the simulations’ results, or in a poor ring alignment, since FWHM varies, on average, by 1.1\% between 975 – 1300 MHz, meaning it is essentially the same for both H and V polarizations at any frequency (and scales similarly across the frequency band). 
On the other hand, a few recently raised issues regarding the test range environment may have caused this behavior. There will be a new series of tests for the second prototype horn and cross-polarization rejection will then be carefully addressed.

On average, the response of the first sidelobe is 25\,dB lower than the main lobe, in agreement with the simulations. These levels of beam sidelobes are similar to those obtained by the VLA L-band horn. The simulated beam pattern in Figure \ref{fig:simshorns} is better than the measured one by about 5\,dB in the second sidelobe, which may be due to the horn support structure, and as such we are investigating ways in which we can improve the measurements and manufacturing process in order to get the horn performance to better match the theoretical value. 

\section{Polariser}
\label{sec:polariser}

As described in Section 2, it is desirable to make observations using circular polarisation. In this section we describe the transition required to connect the circular horn throat to the square input to the polariser, followed by a description of the polarization design itself. Finally, laboratory measurements of the performance are presented.

\subsection{The circular to square transition}
\label{sec:Transition}

\begin{figure}[tb]
   \centering
   \includegraphics[width=0.54\textwidth]{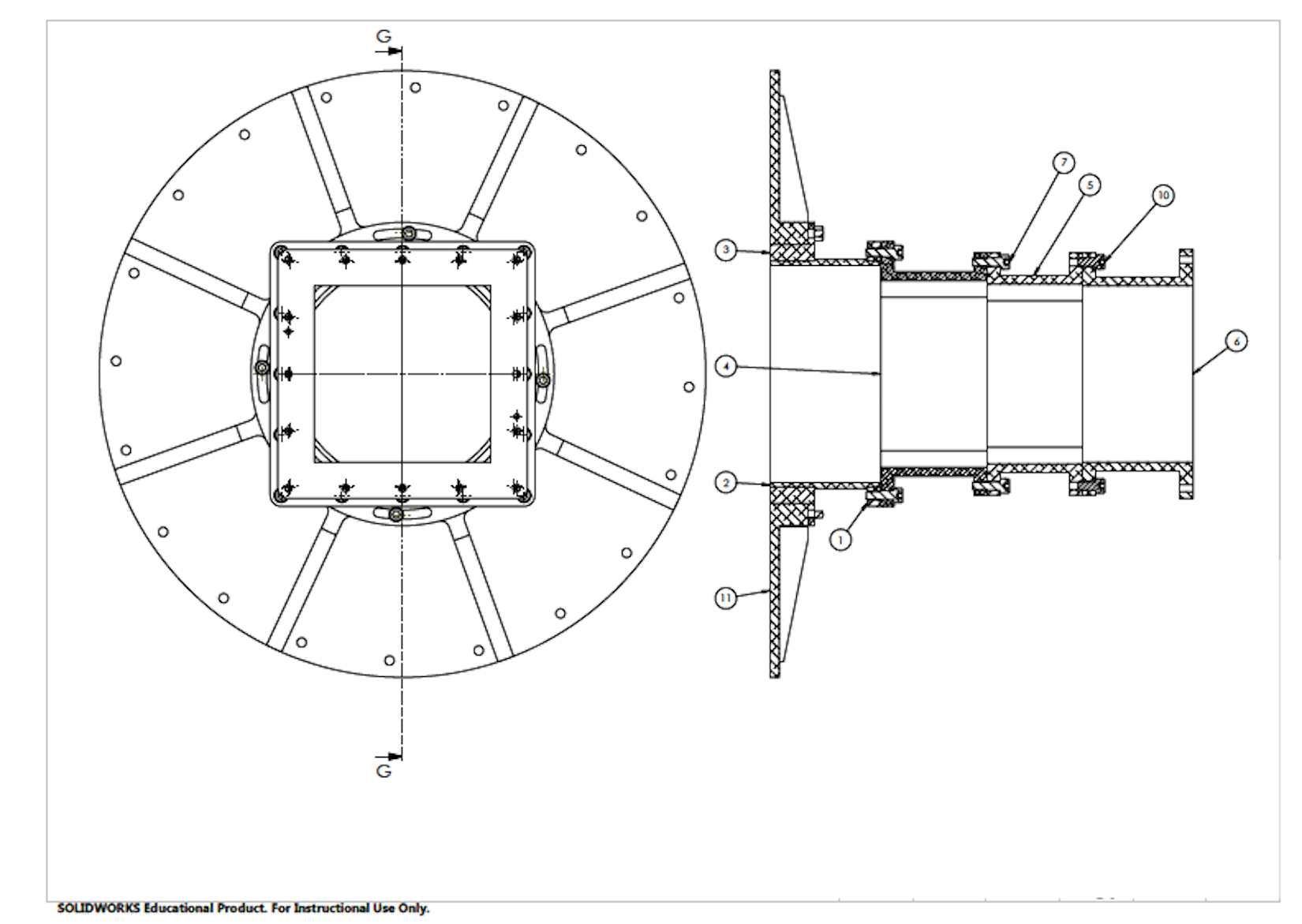}
\includegraphics[width=0.45\textwidth,trim=4cm 4cm 19cm 4cm, clip]{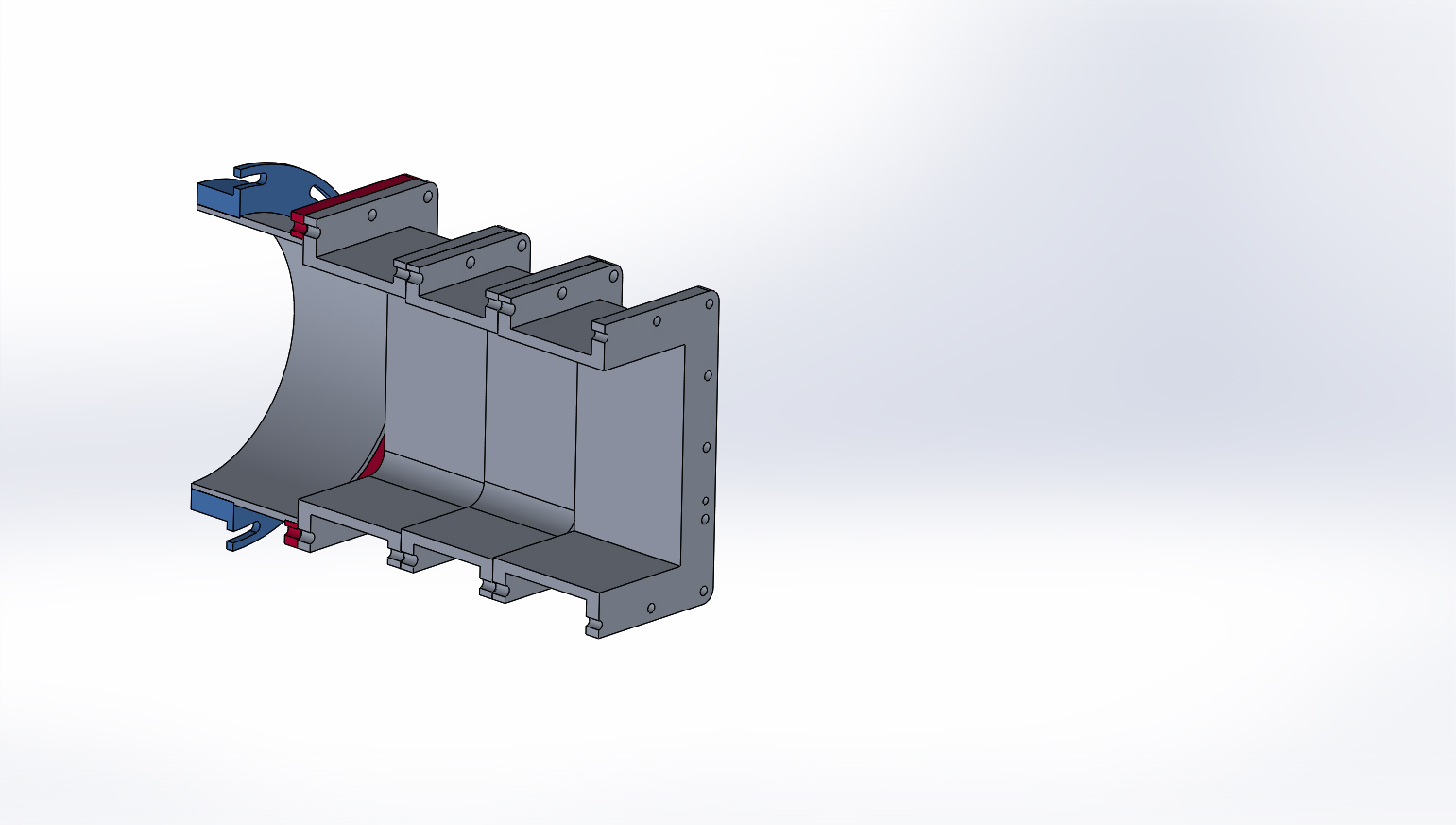}
   \includegraphics[width=0.75\textwidth]{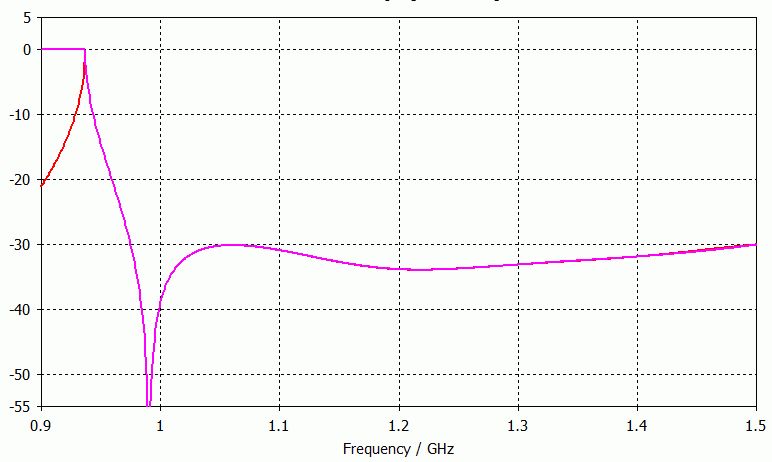}
\caption{Top left: Circular to square transition (front and side view) drawing, used to attach the horn to a polariser. Top right: Circular to square transition, exploded view. Bottom: The predicted return loss of the transition.}
   \label{fig:circ2sqr}
\end{figure}

The termination of the horn is circular but the input to the polariser is a square waveguide so a transition is required. Initially a transformer that gradually merged the circular to square profile was evaluated but in the end, a two-step design was found to give superior performance with a shorter overall length. The total length was under 200\,mm. The design and the predicted return loss are shown in Figure \ref{fig:circ2sqr}. The return loss should be better than $30$\,dB across the band of interest. 



\subsection{Polariser design}

\begin{figure}[tb]
   \centering
   \includegraphics[width=0.75\textwidth,trim=2.5cm 3.5cm 10cm 3.5cm, clip]{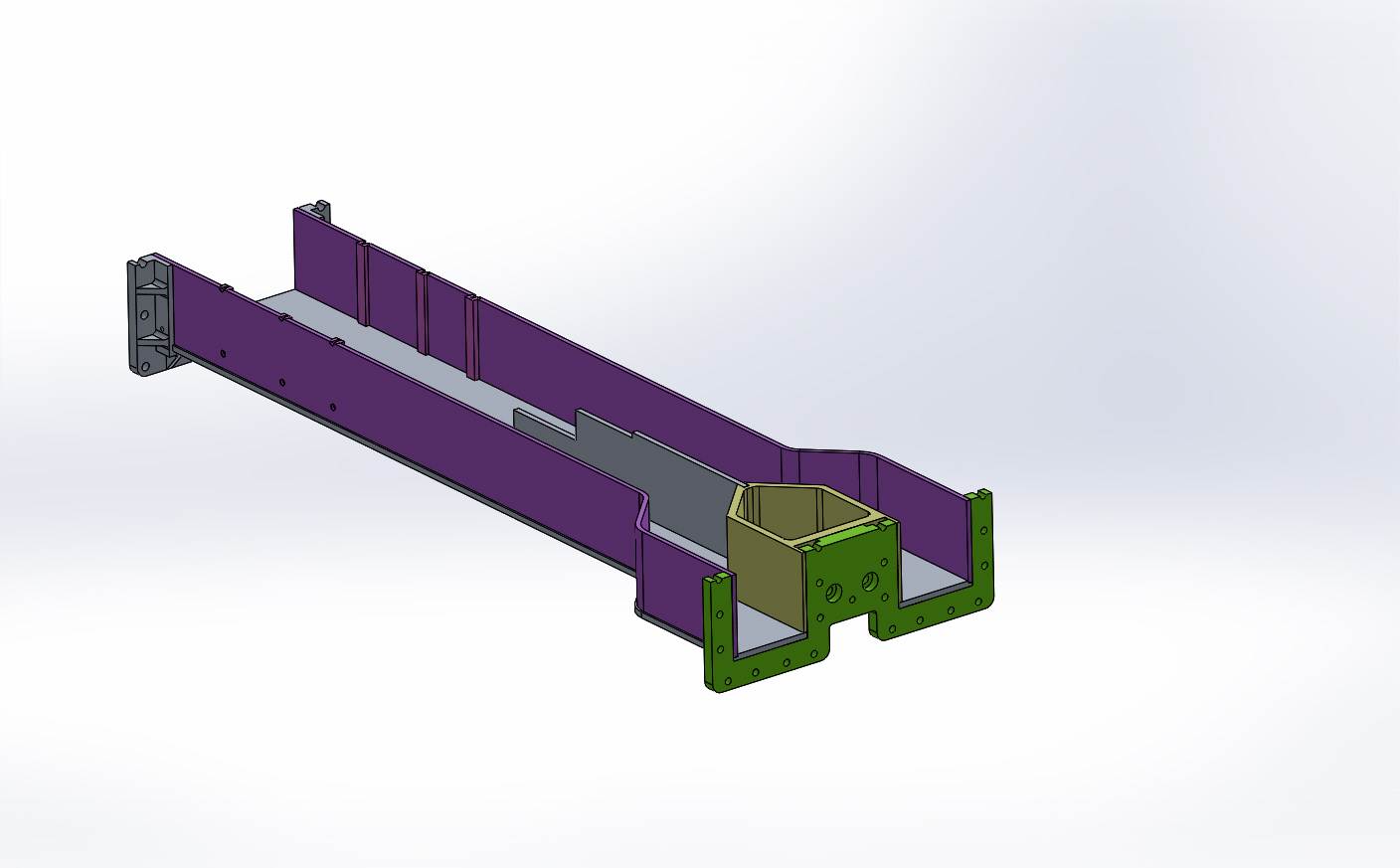}
   \includegraphics[width=0.75\textwidth]{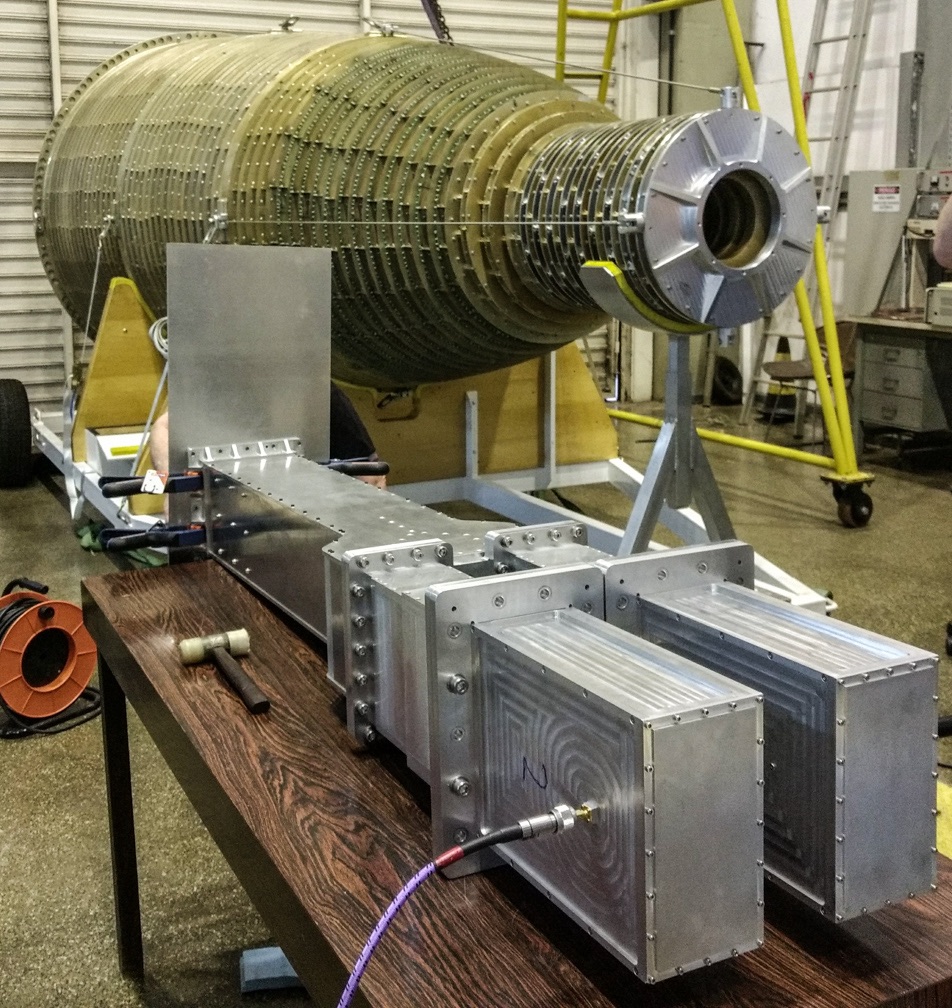}
   \caption{Left: 3D rendering of the internal part of the polariser, showing the septum in detail. Right: Rear view of the transitions, polariser, with the shorting plate at the end of the polarizer square opening, and the horn} 
   \label{fig:polariser2}
\end{figure}

The polariser was designed in the UK by Phase 2 Microwave and manufactured in Brazil by Metalcard. The polariser needs to be able to take the two incoming hands of circular polarisation from the horn and convert them to linear with high isolation between them. The output should be in a rectangular guide because the next component in the receiver chain is a waveguide magic-tee. While specialised waveguide sizes could be used it was decided to make the output in waveguide WG5 (WR770) to simplify connection to the magic-tee and to allow for the use of standard test equipment. Figure \ref{fig:polariser2} shows the 3D rendering of the design.

The relative bandwidth is moderately high at 25\% and the use of a stepped septum polariser is the normal solution for this sort of bandwidth. Since a circular guide version has a narrower bandwidth than the square guide version, the latter was adopted. Also, designs in the square guide were already available at higher frequencies so these could be adapted. The disadvantage is that a circular to square transformer is required as discussed above. The idea of a septum polariser design was originated by \cite{Chen1973}.

A septum with four steps was chosen after a five-step septum was investigated but showed no real benefit in performance. Analytical design is not available so software optimisation of the septum step lengths was required. The design consists of four parts. The septum polariser is followed by a ``joggle'' to separate the output waveguides and two transitions from the special waveguide size to standard WG5. Preceding the septum is a phasing section. The dimension of the square guide was chosen to be 160\,mm. For this, the transverse electromagnetic propagation mode (TE01) cutoff is 937\,MHz, i.e., not too near the 980\,MHz at the bottom of the BINGO band. The next modes are the TE11/TM11 modes at 1325\,MHz. The mode matching Microwave Wizard (from Mician) software was used to optimise the design. The predicted return loss over the band is better than $20$\,dB with 25\,dB isolation between orthogonal polarisations. Figure \ref{fig:polariser2} shows the manufactured hardware. 

To form a pure circular wave we need two linear waves at exactly 90 degrees to each other. It was found that this was very difficult to achieve at the top end of the band and in the end, three pairs of irises, parallel to the septum were used between the septum and the horn which allowed the phase to be improved by 5 degrees to be within $-6$ to $+2.5$ degrees across the required band.

\subsection{Perfomance testing}

%

\begin{figure}[tb]
   \centering
   \includegraphics[width=0.48\textwidth]{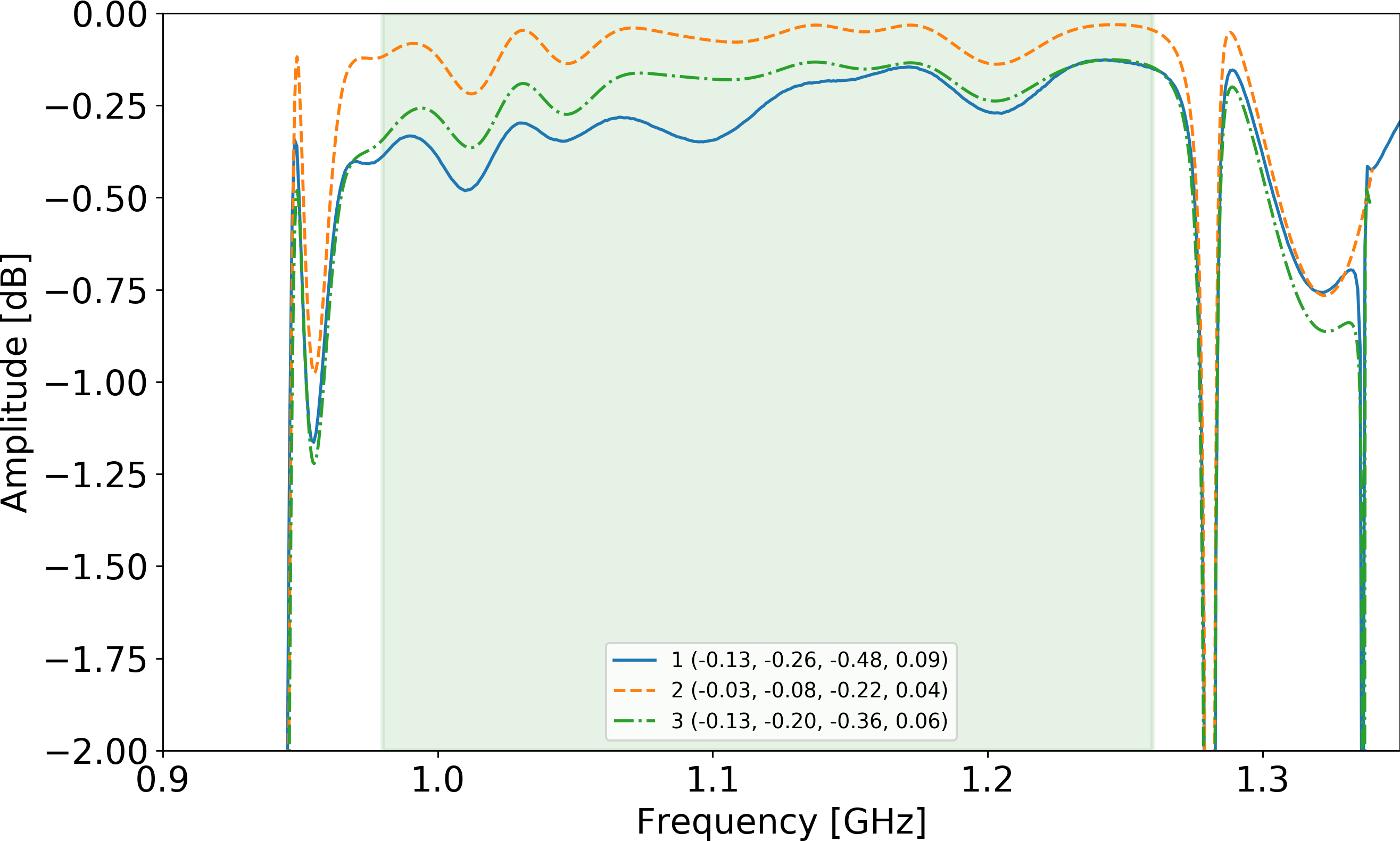}
   \includegraphics[width=0.48\textwidth]{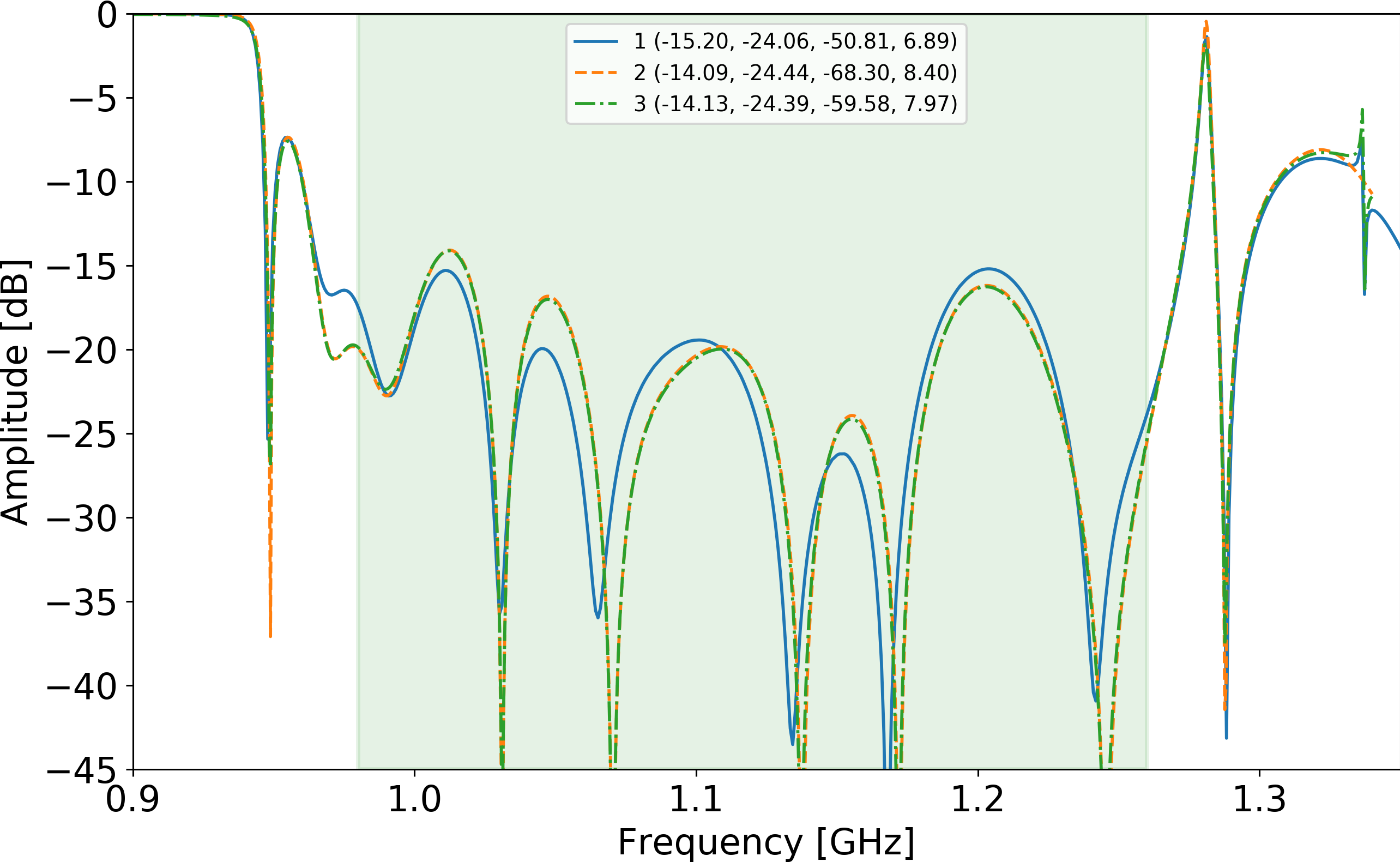}
   \includegraphics[width=0.48\textwidth]{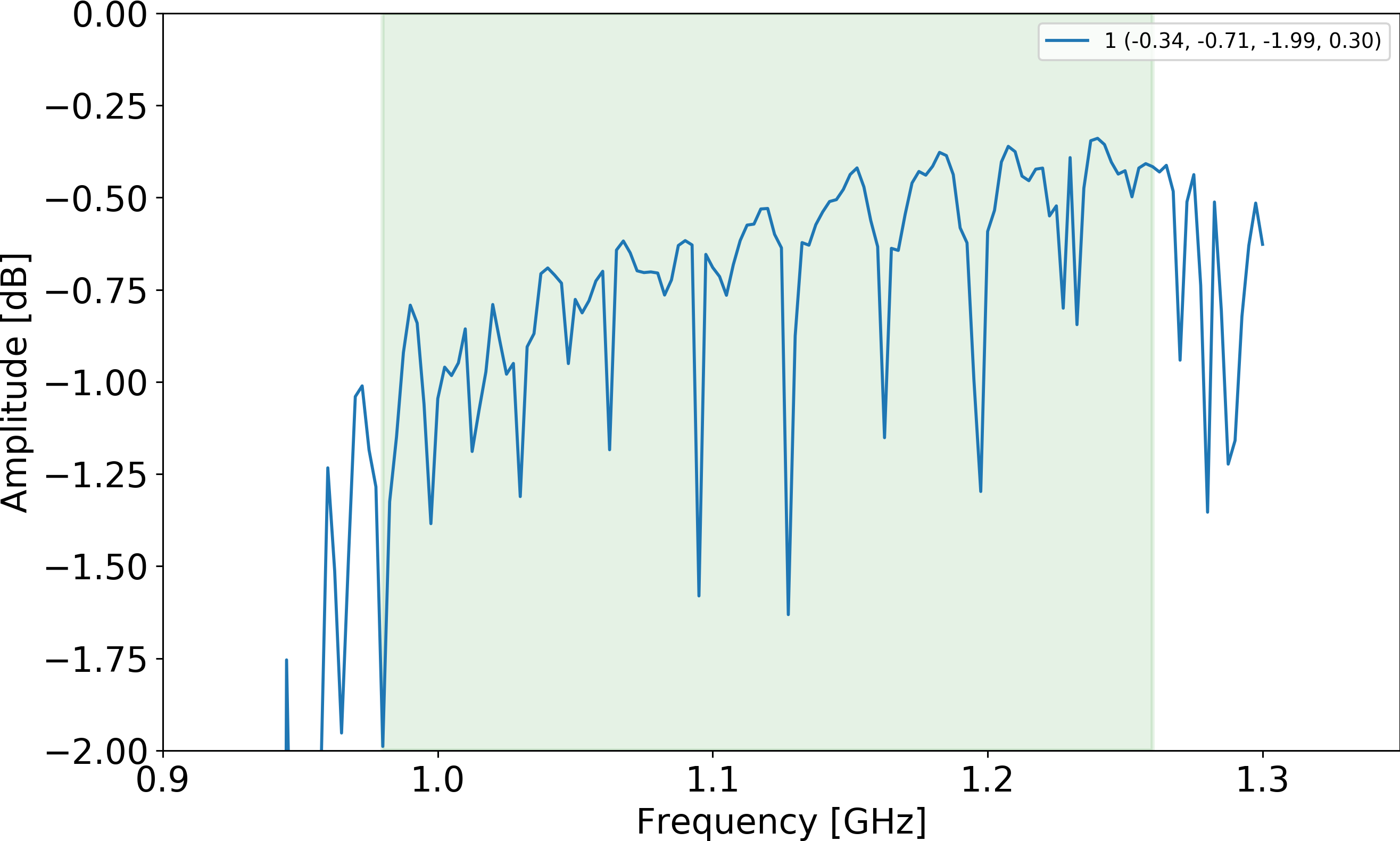}
   \includegraphics[width=0.48\textwidth]{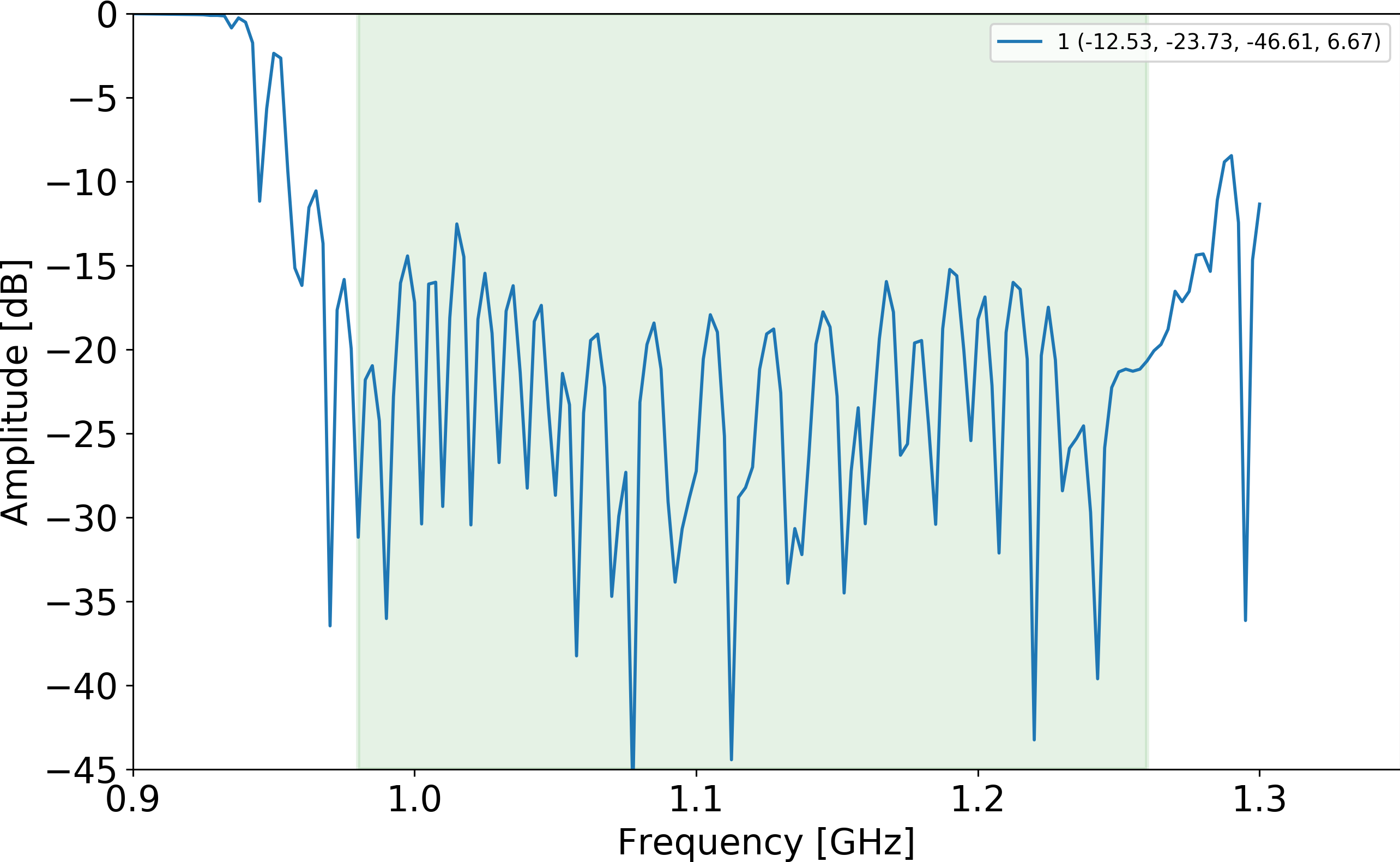}
   \caption{Top: Measurement of the polariser only; the solid blue line is the measurement, while the orange and green dashed lines show simulations with different levels of loss. Left: insertion loss. Right: return loss. Bottom: measurements of the polariser mounted on the horn. Left: insertion loss. Right: return loss}
   \label{fig:pol_meas}
\end{figure}

Measurement of a polariser is difficult and several setups are typically needed to capture all the information. We did not have the benefit of having any waveguide loads, nor two polarisers to perform back to back measurements, which further restricted our options. We decided to proceed as follows: the square polariser port was shorted using an aluminium plate and the polariser was measured between the two rectangular ports. Power entering at one port should be totally reflected at the shorting plate into the other port. The two way loss and both return losses were measured. The results were then compared with theory and a very close match was found for return loss. Measurements for both return losses were almost identical and only one trace is shown. There was a greater deviation for insertion loss, which was higher than desired at $-0.12$\,dB on average across the band (Figure \ref{fig:pol_meas}). We have identified places where fixing the different parts of the polariser together can be improved and it is expected that by doing this we will be able to reduce the loss in subsequent examples.



The combined insertion loss of the horn and polariser were also measured, and are shown in Figure~\ref{fig:pol_meas}. The average insertion loss across the band is $-0.35$\,dB, with a standard deviation of $|0.3|$\,dB, without removing the spikes caused by standing waves. The return loss is $24$\,dB on average across the band.

\section{Summary}
\label{sec:summary}

The return and insertion losses for horn, polariser and transitions call tests are summarised in Table \ref{tab:horn_summary}. The beam response is summarized in Table \ref{tab:beam_measurements}. Altogether, the performance of this horn plus front-end prototype should add approximately 20\,K to the BINGO noise budget.

In addition, within the frequency range 980--1260\,MHz (nominal band) the horn's directivity varies from 22.7\,dB to 25.3\,dB. The difference in FWHM for both polarizations is at most $0.31^{\circ}$ at 1\,GHz, remaining under $0.25^{\circ}$ degrees everywhere else.  The front to back lobe rejection changes within the nominal band. The worst measurements are 48.39\,dB at 1\,GHz (horizontal polarization) and 51.92\,dB at 975\,MHz (vertical polarization). The minimum rejection for both polarizations is around 19.8\,dB at 1150\,MHz and is very likely due to a specific contamination during the measurements. This point will be verified when testing the first inspection unit. Finally, the first sidelobes are located within $20^{\circ}$--$30^{\circ}$ and have amplitudes $-$25\,dB lower than the main lobe.


\begin{table}[htb]
\caption{Horn measurements - summary}
\scriptsize
\centering
\begin{tabular}{@{}c|cc|cc@{}}
\hline
  \multicolumn{1}{c}{Unit} &
  \multicolumn{2}{c}{Insertion loss} &
  \multicolumn{2}{c}{Return loss}  \\
\hline
            & measured              & predicted & measured  & predicted \\
\hline
Horn        &  $-$0.14\phantom{0}   &  --       & 26.50  & 28\phantom{.00} \\
Polariser   &  $-$0.12\phantom{0}   & $-$0.08   & 24.06  & 24.44    \\
WG5 to coax &  $-$0.075             & --        & $> 30$ & --   \\	
\hline
\end{tabular}
\label{tab:horn_summary}
\end{table}

\section{Conclusions and future work}
\label{sec:conclusions}

We have successfully built a large prototype corrugated horn for the BINGO telescope, which achieves low insertion loss (around 0.15\,dB), good return loss (larger than $20$\,dB within the band), and with a clean beam with low sidelobes lower than 27\,dB within the band). The prototype horn will undergo further testing, including use for early astronomical observations at the BINGO site.

A second prototype is under construction, with improvements made to the construction process, such as the addition of a fourth roller to the bending machine, modified aluminium profiles with reinforced corners to reduce tearing during bending. These fabrication adjustments are intended to reduce the amount of human intervention involved in the horn production, which is a major contribution to the final cost when we proceed to the production of the full batch of around 50 horns in the near future. In addition to the mechanical improvements, refinements in the construction of both the horn and polariser are being investigated. Our expectation is that these refinements will improve the electromagnetic performance of the production units, decreasing their contribution to the system temperature.

\begin{acknowledgements}

We thank Jeff Peterson for useful conversations about the horn testing and to Martin Bluck for careful reading of the text. BINGO is funded by FAPESP (\saopaulo\ State Agency for Reasearch Support) through project \mbox{2014/07885-0}. C.A.W. acknowledges support from CNPq through grant \mbox{313597/2014-6}. We specially thank L. Reis, R. Ara\'ujo and G. Kawassaki, from LIT Antenna Testing Facility, for their superb work. M.P. acknowledges funding from a FAPESP Young Investigator fellowship, grant 2015/19936-1.  E.A. wishes to thank FAPESP and CNPq for several grants. V. L. acknowledges the postdoctoral FAPESP grant 2018/02026-0. E. M. acknowledges a Ph.D. CAPES fellowship. F. V. acknowledges a CAPES M.Sc. fellowship.  T.V. acknowledges CNPq Grant 308876/2014-8. Data reduction was performed using Python and the {\sc Matplotlib}, {\sc Numpy} and {\sc Scipy} packages.  
This a preprint of an article published in Experimental Astronomy. The final authenticated version is available online at: https://doi.org/10.1007/s10686-020-09666-9.
\end{acknowledgements}

%
\section*{Conflict of interest}
The authors declare that they have no conflicts of interest.

\bibliographystyle{spphys}
\bibliography{refs}

\end{document}